\documentstyle{mn}

\title[Ly$\alpha$ Imaging of a $z=2.3$ Galaxy] 
{Lyman-${\bmath \alpha}$ Imaging of a Very Luminous ${\bmath z=2.3}
$ Starburst Galaxy with WFPC2\footnote}

\author[N. Roche, J.Lowenthal and B. Woodgate]{Nathan 
Roche$^{1,3}$, James Lowenthal$^{1,4}$ and Bruce Woodgate$^{2,5}$\\
$^1$Department of Physics and Astronomy,
      University of Massachusetts,
      Box 34525,
      Amherst MA 01003,
      USA.\\
$^2$NASA Goddard Space Flight Center, Laboratory for Astronomy and Solar Physics, Code 681, Greenbelt MD 20771, USA.\\
{$^3$ \verb"ndr@wigeon.astro.umass.edu"}\hspace{8mm}
{$^4$ \verb"james@velo.astro.umass.edu"}\hspace{8mm}
{$^5$ \verb"woodgate@s2.gsfc.nasa.gov"}\hspace{8mm}   
}
\input{psfig.sty}

\begin{document}
\maketitle
 
\begin{abstract}
We investigate the $\rm Ly\alpha$ and UV continuum morphology of one of the most luminous known Lyman $\alpha$ emitting galaxies (the `Coup Fourr\'e Galaxy'), associated with a $z=2.3$ damped $\rm Ly\alpha$ absorption system in the spectrum of the QSO PHL 957
 (Lowenthal et al. 1991). The galaxy is observed with the HST WFPC2,  through a narrow filter (F410M) corresponding to
rest-frame $\rm Ly\alpha$ for a total exposure time of  41.2 ksec, plus shorter exposures in F555W and F814W.

In all three passbands, the galaxy is resolved into a close ($\sim 0.35$ arcsec) pair of two components, CFgA and CFgB, both of which are extended and elongated.
  The profile of CFgA is consistent with an exponential disk of similar scale-length in $\rm Ly\alpha$ ($r_{exp}=0.23$ arcsec) and continuum ($r_{exp}=0.20$ arcsec), and  no evidence of a central point source. In contrast, CFgB is closer to a bulge profile.
       We find that CFgA has by far the  higher ratio of $\rm Ly\alpha$ to continuum flux, and from the observed colours estimate rest-frame equivalent widths of $W(\rm Ly \alpha) =151\pm 16 \AA$ for CFgA and $\rm 33\pm 13 \AA$ for CFgB. 

From the F814W and F555W magnitudes we estimate rest-frame blue-band absolute magnitudes (for $H_0=50$ km $\rm s^{-1}Mpc^{-1}$ and $q_0=0.05$) of  -23.12 for CFgA and -23.24 for CFgB, significantly brighter than local galaxies of the same size. CFgA shows a remarkable 3.9 magnitudes of surface brightness enhancement relative to local spirals.  This object appears to be at the upper limit of both the range of 
surface brightness evolution observed in $z>2$ galaxies and the range of $W(\rm Ly \alpha)$ in any star-forming galaxy. We speculate that its extreme surface brightness results from a very luminous starburst ($\sim 200 M_{\odot}\rm yr^{-1}$), triggered by the merger of the two components, and the high $W(\rm Ly \alpha)$ from a brief phase of the starburst in which most $\rm Ly\alpha$ photons can escape, as predicted in the models of Tenorio-Tagle et al. (1999).
 
We also investigate the F410M image of the QSO PHL 957. Subtraction of a normalized point-spead function leaves no significant residuals -- the QSO is consistent with a pure point source and we do not detect either the host galaxy or the damped $\rm Ly\alpha$ absorbing galaxy. 
 
We search for other galaxies with strong $\rm Ly\alpha$  emission at  $z\sim 2.3$--2.4, selecting these by a colour $(m_{410}-V_{555})_{AB}<-0.2$. Eight candidate $\rm Ly\alpha$ sources, all fainter than the Coup Fourr\'e galaxy, are identified in our field. One is a point-source and may be an AGN; the others are of similar size to the Coup Fourr\'e Galaxy but lower surface brightness, knotty and asymmetric. They appear typical of Lyman break galaxies but with colours indicating $W(\rm Ly\alpha)\sim 100\rm \AA$.  
\end{abstract}
 
\begin{keywords}
Galaxies -- evolution: galaxies -- starburst: ultraviolet -- galaxies
\end{keywords}
{\thefootnote Based on observations with the NASA/ESA Hubble Space Telescope obtained at the
Space Telescope Science Institute, which is operated by the Association of Universities for Research in Astronomy, Incorporated, under NASA contract NAS5-26555.}
\section{Introduction}

Clouds of neutral hydrogen detected as Lyman $\alpha$ ($\rm Ly\alpha$, $1216\rm \AA$) absorption lines in the spectra of
distant QSOs sample gas in the early universe, without bias towards
luminosity or the presence of stars.  They are therefore useful as
sensitive indicators of the distribution, metallicity, and ionization
of gas at redshifts as high as the most distant QSOs ($z>5$),
and they complement galaxies selected by emission, such as Lyman break
galaxies.  In particular, damped  $\rm Ly\alpha$ absorption systems (hereafter DLAs), with neutral hydrogen
column densities $N_{\rm HI} > 10^{20}\rm cm^{-2}$, although rare, contain most of the
HI in the universe at high redshift, and have been postulated to
correspond to the sites of young, possibly primeval galaxies, and of cluster formation (e.g.,
Wolfe 1993).  Despite intense efforts to
image dozens of DLA clouds directly in optical continuum or
emission lines, only a few systems at $z>2$ have yielded any
associated emission (e.g., Djorgovski et al 1996;
Francis et al. 1996; M\o ller and Warren 1998; Fynbo, M\o ller,
and Warren 1999; Leibundgut and Robertson 1999)
 -- note that the latter three cases
have $z_{abs} \simeq z_{em}$, so the cloud is presumably influenced by
the QSO close by.

The QSO PHL 957, at $z=2.681$, shows the spectral signature of damped 
$\rm Ly\alpha$ 
absorption from a neutral hydrogen cloud ($N_{\rm HI}=2.5\times 10^{21}\rm cm^{-2}$) at $z=2.309$.
 Meyer and Roth (1990) detected the NiII, CrII and ZnII absorption lines from the cloud and from the relative line strengths concluded  that this DLA has a metallicity (Zn/H) only 4 per cent solar, and a dust-to-gas ratio
only 3 per cent Galactic. 
This very low metallicity and dust content implies that any $\rm Ly \alpha$ emission from within, or associated with, the system, e.g. from star-forming galaxies, might suffer very little dust obscuration, allowing its direct observation.
  
 To investigate this possibility, the PHL 957 field was observed (Lowenthal et al. 1991) at the KPNO 4m telescope through a Fabry-Perot filter tuned to $4023\pm 28\rm \AA$ (rest-frame $\rm Ly\alpha$). Significant  emission was detected from a small ($<3$ arcsec) source 48 arcsec from PHL 957. Spectroscopic investigation at the Multiple-Mirror telescope (MMT) revealed a faint ($V\simeq 23.6$) continuum with a strong $\rm Ly\alpha$ line of flux $5.6\times 10^{-16}$ ergs $\rm s^{-1}cm^{-2}$, or rest-frame equivalent width $W(\rm Ly\alpha)\simeq 140\rm \AA$. The line gave a 
redshift $z=2.3128\pm 0.0004$  -- indicating a velocity $346\pm 20$ km $\rm s^{-1}$ relative to the absorbing cloud -- and showed Doppler broadening corresponding to ${\rm FWHM}\sim 600$ km $\rm s^{-1}$. 
The $\rm Ly\alpha$ source was therefore a galaxy, later named the Coup Fourr\'e galaxy (hereafter, CFg), separate from but associated with the absorbing cloud.
The CFg spectrum also showed less luminous CIV($\rm 1549\AA$) and HeII($\rm 1640\AA$) emission lines, which might indicate the presence of an Active Galactic Nucleus (AGN), but could also be produced by Wolf-Rayet stars.

Observing CFg in the near infra-red, Hu et al. (1993) obtained a marginal detection of $\rm H\alpha$ emission, confirmed at higher significance by 
Bunker et al. (1995, 1999). 
 Wolfe et al. (1994) confirmed that $\rm[O/H]<-0.97$ solar for the damped absorption cloud, indicating that the system's metallicity is sufficiently low that there is at least a possibility for any $\rm Ly\alpha$ photons to escape (Charlot and Fall 1993).
However, these
ground-based observations were not able to reveal either the spatial distribution of the $\rm Ly\alpha$ emission or the morphology of the underlying galaxy, and it remained unclear whether the $\rm Ly\alpha$ flux was produced by an AGN, by more extended star-formation, or some combination of the two.

A number of  other $\rm Ly\alpha$ emitting galaxies have now been identified
at high redshifts.  One of the most luminous, at $z=3.428$, was observed in the $V$-band  (Giavalisco et al. 1995) with the pre-refurbishment HST WFPC, 
and  found to have 
a compact ($r_{hl}=0.09\pm 0.02$ arcsec) bulge-profile morphology.
M\o ller and Warren (1993) investigated three $\rm Ly\alpha$ emitting galaxies at $z=2.81$, associated with the DLA towards QSO PKS0528-250. HST WFPC2 imaging
in broad-band $B$ and $I$ revealed these also to be compact,
$r_{hl}\simeq 0.1$ arcsec, in the rest-frame UV continuum (one consisted of two components separated by 0.3 arcsec), although  ground-based narrow-band imaging suggested that one source was more extended in $\rm Ly\alpha$ ($r_{hl}\simeq 0.51$ arcsec).

The obvious next step in the study of this type of galaxy is deep space-based,
narrow-band imaging at the  redshifted $\rm Ly\alpha$ wavelength, combined with  broad-band imaging. This will 
enable the $\rm Ly\alpha$ and continuum structures to be investigated and compared at sub-kpc scales, revealing whether the $\rm Ly\alpha$ emission is from a point-source or extended region, and if extended whether it traces the distribution of stars, and indicating  the morphology (e.g. disk, bulge or merger) of the underlying galaxy. Fortunately, one of the WFPC2 filters, the medium-width F410M, covers the desired wavelength, but a very long exposure time is needed.
We were allocated a total of 21 orbits of WFPC2 time for the detailed study of CFg, of which 16 orbits were dedicated to acquiring the $\rm Ly\alpha$  image. 

 In addition, any other strong
$\rm Ly\alpha$ sources on our field at redshifts close to CFg will be distinguishable from other galaxies by an enhanced ratio of F410M to broad-band  flux.  
Pascarelle, Windhorst and Keel (1998) applied this technique to a field centred on the $z=2.4$ radio galaxy  53W002, plus three randomly chosen fields, which were imaged with WFPC2 in F410M, F450W and F814W. In the 4 fields, a total of 35 galaxies were selected as probable $\rm Ly\alpha$ emitters on the basis of blue $m_{410}-B_{450}$ colours, with the greatest number (17) in the 53W002 field. 
Similarly, a narrow-band $5007\rm \AA$ survey of Kudritzki et al. (2000) revealed 9 $\rm Ly\alpha$ emitters at $z=3.1$.
However, not all high-redshift galaxies are $\rm Ly\alpha$ emitters --
Steidel et al. (1999) performed
ground-based $\rm Ly\alpha$ imaging of a field with many identified Lyman break
galaxies at $2.7\leq z \leq 3.4$ and estimated that only $\sim 20$--25 per cent of  these have $W(\rm Ly\alpha)>20 \AA$.   

Lastly, 
subtracting a normalized point-spread function from the image of a QSO may reveal the host galaxy (e.g. McLure et al. 1999), although this is more difficult for high-redshift QSOs (e.g. Lowenthal et al. 1995), and/or show a galaxy at the centre of the DLA system. We attempt this with the F410M image of  PHL 957.

Section 2 describes the observational data and its reduction. In Section 3 we present contour plots and radial profiles of CFg, and estimate the scalelength and morphological type of this galaxy. In Section 4 we investigate the F410M image of PHL 957. In Section 5, model galaxy spectra are used to interpret our  results in terms of $W(\rm Ly \alpha)$, surface brightness evolution etc. 
In Section 6, colour-selection is used to identify other possible  $\rm Ly\alpha$ emitters. In Section 7 we discuss
our results concerning the morphology, environment and evolution of CFg, the
$\rm Ly\alpha$ properties of high-redshift galaxies in general, and possible future observations.

\section{Observations}
\subsection{Data}
With the HST WFPC2, a field centred on RA $01^h03^m8.4^s$, Dec. $+13:16:40$ (equinox 2000.0)  - to include both the CFg and the QSO PHL 957 -- 
was observed through 3 filters.
The F410M observations consist of 32 ($4\times 1200$ sec and $28\times$ 1300 sec) exposures (41.2 ksec total) obtained during
16 HST orbits on 3 and 5 December 1998. The F555W and F814W (wide-band) observations were made during 5 HST orbits on 16 January 1999, and consist of $2\times 1200$ sec and 
$4\times 1300$ sec exposures (7.6 ksec total) in F555W (hereafter $V_{555}$) and $4\times 1300$ sec (5.2 ksec total) in F814W (hereafter $I_{814}$, the standard WFPC2 $I$-band). 

\subsection{Filters}
Figure 1 shows the response curves for the three, non-overlapping filters of our observations, together with a low-resolution ($\rm FWHM\simeq 24\AA$) CFg spectrum obtained in a 3423 sec exposure with the MMT (Lowenthal et al. 1991).
Figure 2 shows the $\rm Ly\alpha$ line on a medium-resolution ($\rm FWHM\simeq 2.6\AA$) 3350 sec MMT spectrum, with the F410M curve overplotted.  Evidently, the strong $\rm Ly\alpha$ line will dominate the flux observed through F410M, while the broad-band filters sample a relatively featureless UV continuum. 
 At the CFg redshift of $z=2.3128$, the FWHM ranges of the filters are 1206--$1261\rm \AA$ for F410M, 1348--$1828\rm \AA$ for F555W and 2137--$2909\rm \AA$ for F814W. The $\rm Ly\alpha$ line will lie within the FWHM of F410M for sources at $2.286<z<2.436$. 
\begin{figure}
\psfig{file=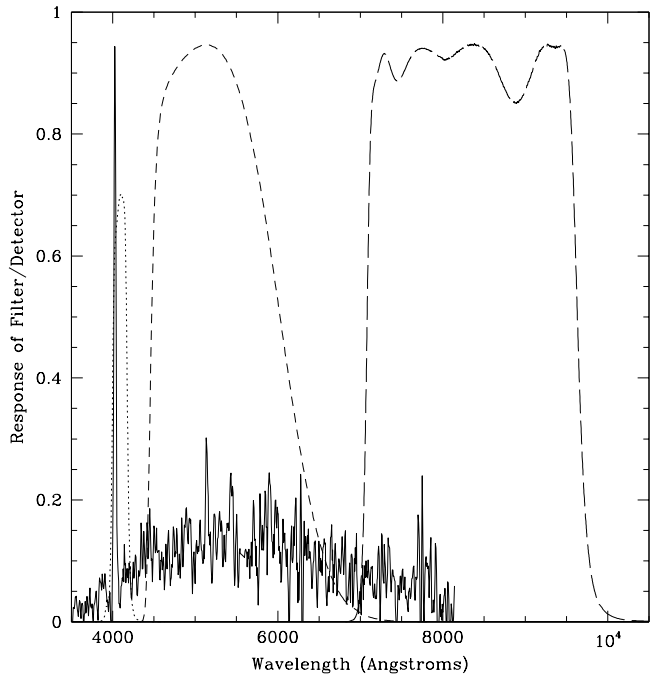,width=95mm}
\caption{Response curves for the F410M (left,dotted), F555W (centre, short-dash) and F814W (right, long-dash) filters, together with the low-resolution MMT spectrum of the CFg (Lowenthal et al. 1991), here shown slightly smoothed.}
\end{figure}
\begin{figure}
\psfig{file=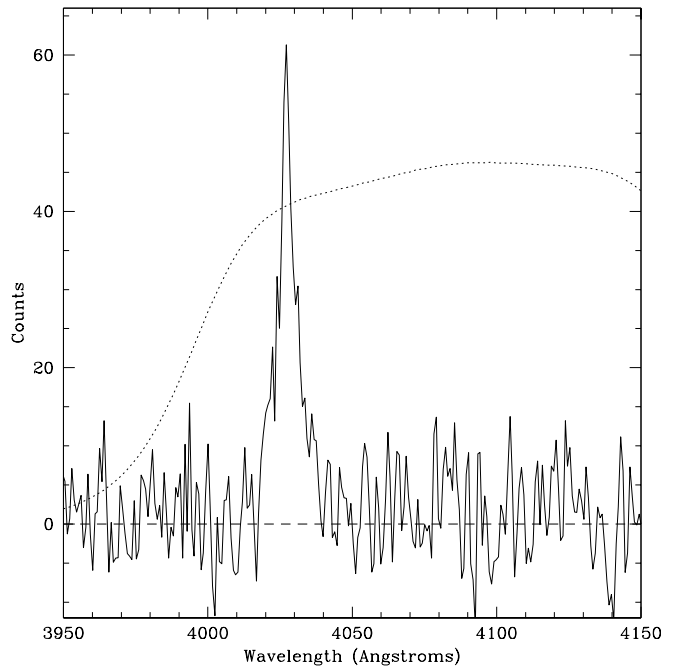,width=95mm}
\caption{Medium-resolution spectrum of the $\rm Ly\alpha$ line of the CFg
(Lowenthal et al. 1991), unsmoothed, with the F410M response curve (dotted).}
\end{figure}

\subsection{Data Reduction}

Data reduction was carried out using the {\sevensize IRAF} package with some additional routines.  We received 
the data, already debiased, flat-fielded and calibrated, for the 42 individual
exposures.

 The first problem was the removal of the large numbers of cosmic rays
from the images. For this purpose we used the `nukecr' {\sevensize IRAF} 
routine developed by Luke Simard specifically for HST data. This generates a median-filter smoothed image from the pair of exposures from each HST orbit, and detects cosmic rays as pixels discrepant from this by more than a pre-set threshold. Pixels within 0.2 arcsec of each cosmic ray detection, in both exposures of the pair, are then replaced by the correponding pixels from the smoothed image.
With a careful choice of thresholds, it was possible to completely remove the majority of visible cosmic rays and reduce the others to a few pixels, with no effect on the photon counts from the real stars and galaxies.

 WFPC2 images consist of 4 chips (c1--c4), each with $800\times 800$ pixels, with the pixel size 0.0996 arcsec on c2--c4 and 0.0455
arcsec on c1 (the PC). The PC data proved to be of poorer signal-to-noise and hereafter we are concerned only with c2--c4. The pixel size of these chips undersamples the point-spread function (PSF), but when stacking a large number of slightly offset exposures, it is possible to regain much of the `lost' resolution by first rebinning all images into a smaller pixel grid.

During the observations, the WFPC2 field centre was shifted by 1 or 2 arcsec between each orbit (dithering) so as to minimize the effects of bad pixels on the stacked images.
These positional offsets were measured to $< 0.01$ arcsec accuracy using the cross-correlation technique of {\sevensize IRAF} `precor', `crossdriz' and `shiftfind'.  
With the {\sevensize IRAF} `drizzle' routine, all exposures (together with their data quality files) were rebinned by a factor 0.42, into  0.0418 arcsec pixels, 
and simultaneously positionally registered using our measured offsets. 

In each passband, the drizzled exposures are then stacked using {\sevensize IRAF} `imcombine'. At each pixel position, counts discrepant by chosen thresholds (approximately $2.5\sigma$) from the median of the stack, and those
 from `bad pixels' as listed in the data quality files, are rejected from the
averaging.  This `sigclip' rejection was effective in removing almost all of the remaining cosmic ray contamination from the combined F410M and F555W data. It was less effective in F814W, where there were only 4 exposures, so the combined F814 image was further cleaned 
 using  {\sevensize IRAF} 'cosmicrays'. Again, we checked that these procedures
 did not remove any of the signal from the real objects. 
 
\subsection{Source Detection and Photometry}

Sources on the combined images were detected and catalogued using the most recent version of SExtractor (Bertin and Arnauts 1996), which begins by fitting a sky background and estimating the sky noise $\sigma_{sky}$. We chose 
a detection criterion of $\geq 2\sigma_{sky}$ above the background in $\geq 16$ contiguous pixels, with a Gaussian 2.5 pixel FWHM filter. The contrast threshold for source deblending was set to a relatively high 0.1.

 Photometric zero-points are derived from the 
`photflam' parameters in the image headers, multiplied by $\lambda_{pivot}^2$
to convert from $F_{\lambda}$ to $F_{\nu}$. Throughout this paper (unless stated otherwise) all magnitudes are given in the AB system, in which the magnitude  for any passband is 

$m_{AB}=-2.5~{\rm log}_{10}~F_{\nu}-48.60$,

 where 
$F_{\nu}$ is flux in units ergs $\rm s^{-1} cm^{-2} Hz^{-1}$. The stacked images are normalized to the mean exposure time of 1287.5, 1266.7 and 1300.0 sec in F410M, F555W and F814W respectively. One count on these images then corresponds to $m_{410}=27.273$, $V_{555}=30.315$ and
$I_{814}=29.864$. 
On the basis of these zero-points the mean sky brightness is  $m_{410}=23.20$,
$V_{555}=21.95$ and $I_{814}=21.49$ mag $\rm arcsec^{-2}$ and our surface brightness detection thresholds ($2\sigma_{sky}$) are  $m_{410}=23.55$,
$V_{555}=24.71$ and $I_{814}=24.17$ mag $\rm arcsec^{-2}$.

SExtractor detected 
a total of 71 sources on the 3 chips of the F410M data, 411 in $V_{555}$ and 491 in $I_{814}$. For each detection, 
SExtractor gives fluxes in circular apertures of fixed diameter (here 2.0 arcsec) and `Kron' fluxes (in elliptical apertures fitted to each object), and converts these to magnitudes using the photometric zero-points.

 \section{The Nature of the Coup Fourr\'{e} Galaxy}
\subsection{General Appearence}
The CFg lies at $z=2.3128$, where for $q_0=0.05$ and $H_0=50h_{50}$ km $\rm s^{-1} 
Mpc^{-1}$ (assumed throughout the paper) , 1.0 arcsec is $12.2 h_{50}^{-1}$ kpc. In F410M
it is detected as one of the brighter sources on chip 3, at a position which the astrometry supplied with the WFPC2 data 
gives as  RA $01^m03^m8.45^s$ Dec. $+13:16:41.42$, close to the spectroscopy position of  RA $01^m03^m8.43^s$ Dec. $+13:16:39.61$ from Lowenthal et al. (1991) (converted to equinox 2000.0).

With SExtractor, the CFg is  a $\sim 14.5\sigma$ significance detection in F410M, with $m_{410}=23.18\pm 0.07$ in a 2 arcsec diameter aperture. It is clearly extended, with a Gaussian FWHM of 0.70 arcsec compared to 0.14 arcsec  for the point-spread function (PSF), and elongated (ellipticity of 0.402). On the broad-band images, it is detected  with aperture magnitudes  $V_{555}=23.80\pm 0.03$ and $I_{814}=23.48\pm 0.03$, and appears
similar in size. 

 To best estimate the colours of the $\rm Ly\alpha$ emitting galaxy, SExtractor is run in a `double-image mode' whereby detection and Kron aperture-fitting are performed in F410M, but fluxes measured from the corresponding pixels of the F555W and F814W images. 
Comparing the Kron magnitudes from this method gives $m_{410}-V_{555}=-0.56\pm 0.08$ and 
$V_{555}-I_{814}=0.36\pm 0.01$, consistent with the colours from the larger circular apertures.
\begin{figure}
\psfig{file=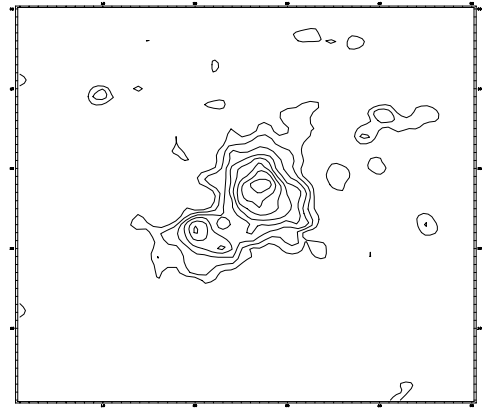,width=95mm}
\caption{ Contour plot of the $2.1\times 2.1$ arcsec area centred on CFg
in F410M, with contours linearly spaced in flux from the $2\sigma_{sky}$ detection threshold to the peak intensity. Plot is oriented N at the top, E at the left.}
\end{figure}
\begin{figure}
\psfig{file=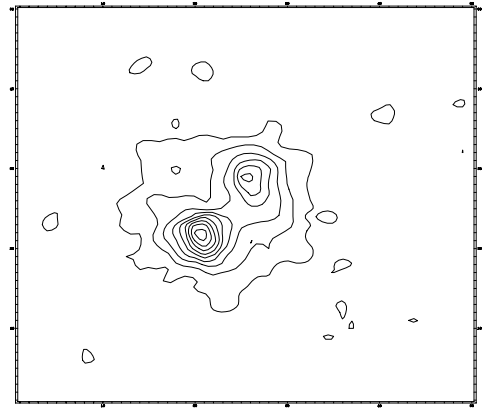,width=95mm}
\caption{As Figure 3, in F555W.}
\end{figure}
\begin{figure}
\psfig{file=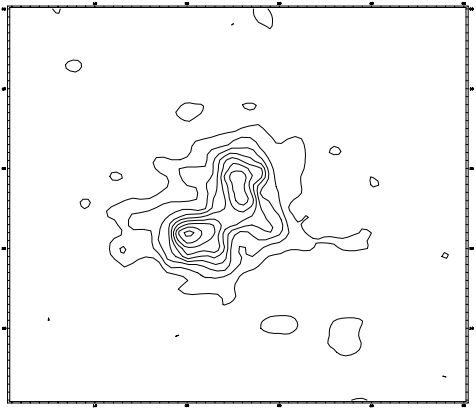,width=95mm}
\caption{As Figure 3, in F814W.}
\end{figure}
Figures 3, 4 and 5 show contour plots of the CFg in the three passbands. On  chip 3 North is $71.7^{\degr}$ anticlockwise of the y-axis, but these plots are rotated to show North at the top and East at the left.
The CFg is seen to consist of two similarly-sized and elongated components within a
common envelope, of which the northwestern 
(hereafter CFgA) is the more prominent in F410M, whereas in $V_{555}$ and $I_{814}$ the southeastern (hereafter CFgB) has the higher surface brightness.
\subsection{Surface Brightness and Colour Profiles}
The profile of CFg is investigated using the {\sevensize IRAF} `isophote' package, which fits the isophotes surrounding a chosen intensity peak with a series of concentric ellipses. We first fit isophotes to the F410M profile, which is peaked at the CFgA nucleus, and show on Figure 6 the mean surface brightness of the F410M isophotes as a function of semi-major axis. 
\begin{figure}
\psfig{file=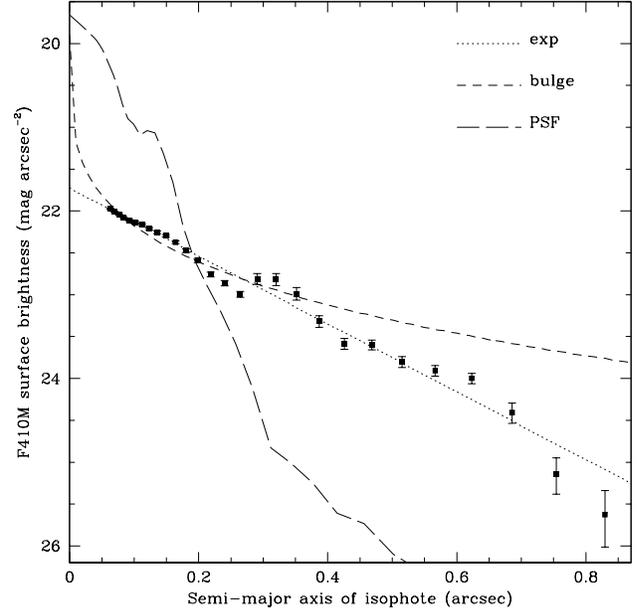,width=89mm}
\caption{F410M surface brightness (above sky background) of isophotes centred on the CFgA nucleus, as a function of semi major axis, with the best-fitting exponential (dotted) and bulge (short-dashed) profiles, and the PSF normalized to the same total flux as CFg (long-dash).}
\end{figure}

 Figure 6 also shows the profile given by applying the same analysis to the F410M PSF at the CFg  position, as simulated using the `Tinytim' package (Krist 1995).
The observed FWHM of the CFg is $\sim 5$ times that of the PSF, so estimation of the galaxy size  from the observed profile should (on the basis of subtracting the FWHM in quadrature) overestimate by only $\sim 2$ per cent. Hence we neglect the effect of the PSF, and quantify the galaxy size by fitting (by the least-squares method) the observed relation of surface brightness ($\mu$, in mag $\rm arcsec^{-2}$) to semi-major axis ($r$) with (i) an exponential (disk) profile
$$\mu=\mu_0~{\rm exp}(-r/r_{exp})$$
and (ii) a bulge (elliptical) profile
$$\mu=\mu^b_0~{\rm exp}(-7.688[r/r_e]^{0.25}).$$ 
\begin{figure}
\psfig{file=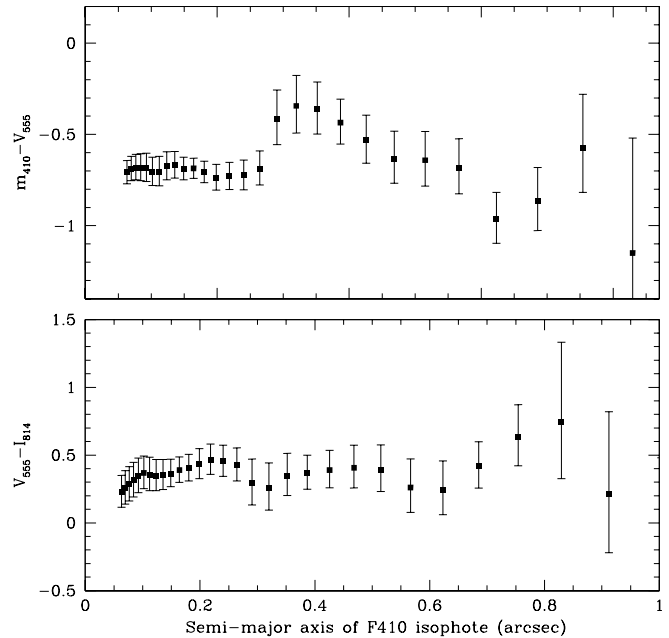,width=91mm}
\caption{Mean $m_{410}-V_{555}$ (above) and $V_{555}-I_{814}$ (below) colours (AB system) on the F410M isophotes of the CFg.}
\end{figure}

 As the closely-spaced isophotes and the  resampled pixels are non-independent, the $\chi^2$ of the fits will be overestimates, and we instead  quantify the goodness of fit in terms of the rms residual in magnitudes. The F410M profile is fitted at 28 isophotes in the range $0.063\leq r\leq 0.83$ arcsec.
\onecolumn
\begin{figure}
\psfig{file=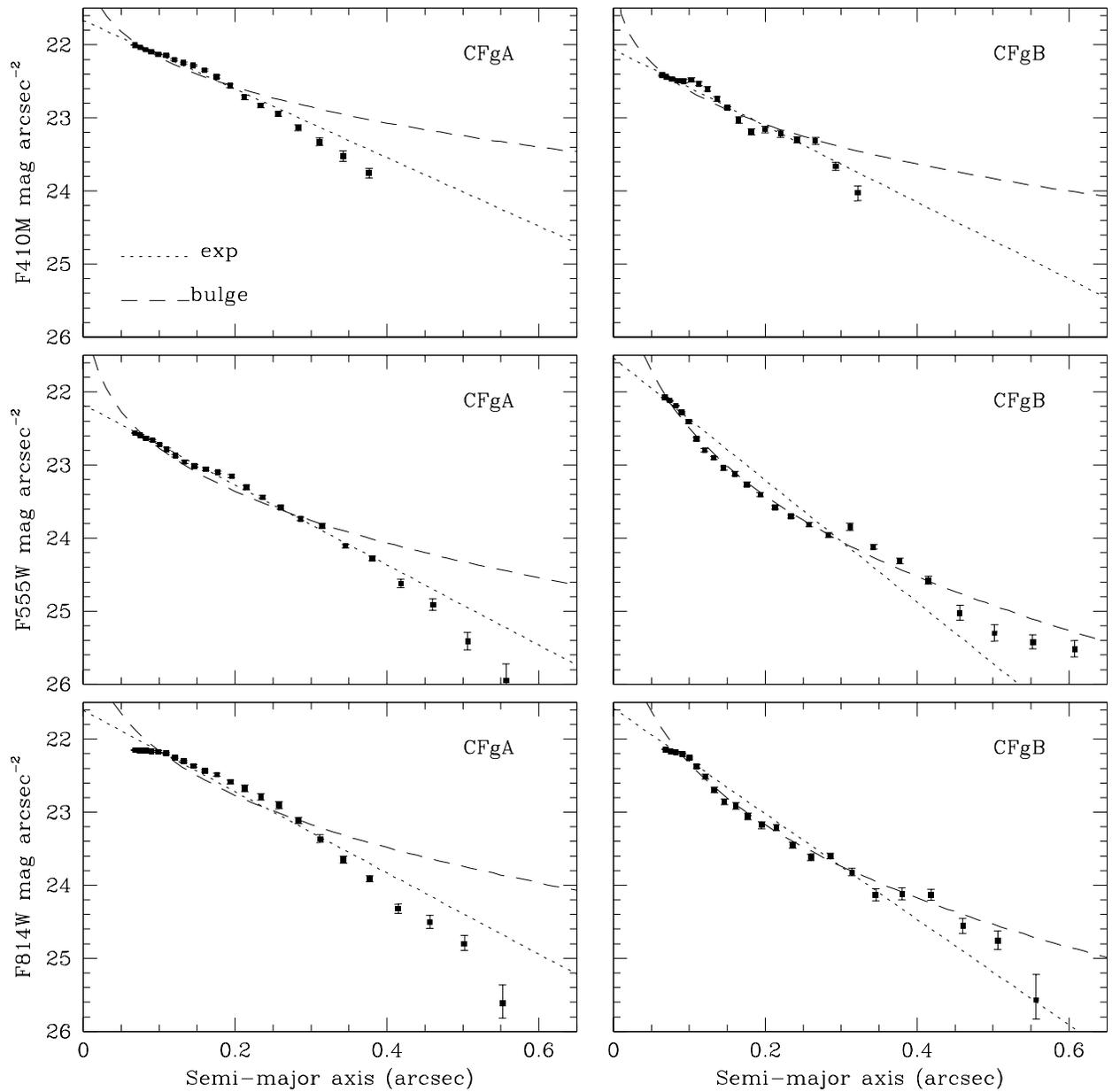,width=180mm}
\caption{Surface brightness as a function of semi-major axis, for isophotes fitted in F410M (top), F555W (mid) and $I$ (lower) to  CFgA (left) and CFgB (right), with the other component masked out. The higher signal-to-noise of the broad-band images allows isophotes to be fitted to larger radii. The dotted and dashed lines show the exponential and bulge profiles best-fitting each observed profile.}
\end{figure}
\twocolumn
The exponential fit gives $\mu_0=21.73$ and $r_{exp}=0.27$ arcsec, with rms
residuals 0.16 mag. The bulge fit gives $\mu^b_0=19.90$ with an implausibly large $r_e$ (as typically results from attempting to fit a disk galaxy  with a bulge profile) of 17.99 arcsec and much larger residuals of 0.48 mag. 

Firstly, the CFg profile is evidently closer to an exponential than to a bulge, although it shows a `bump' at $r\simeq 0.3$ arcsec produced by the fainter (in F410M) CFgB nucleus. 
Secondly, there is
no indication of any central point-source of $\rm Ly\alpha$ emission in excess of the exponential profile. Any point-source would have to be $>2$ mag fainter than the underlying galaxy so as not to produce a visible ($\geq 0.3$ mag) turn-up in the profile at $r<0.15 $ arcsec.  

Using `isophote', we  measure the mean $V_{555}$ and $I_{814}$ surface brightness on the fitted F410M isophotes, to obtain colour profiles (Figure 7).
A colour $m_{410}-V_{555}\simeq -0.7$ is seen at all radii from the CFgA nucleus out to $\simeq 0.27$ arcsec, where the isophotes become  significantly redder  due to the contribution of CFgB.
In contrast, $V_{555}-I_{814}\simeq 0.3$--0.4 over the whole galaxy,  
to $\geq 0.7$ arcsec, and  if anything the CFgB nucleus causes a slight bluening.

The marked bimodality of the surface brightness profile (even greater in the broad-bands), and the uniformity of $m_{410}-V_{555}$ within CFgA compared to the large CFgA/CFgB difference, imply that the CFg
is a system of two distinct components, which might have very different properties, and might be best studied by 
considering  CFgA and CFgB separately.
\subsection{The CFg as a Double System}
Running SExtractor with a lower contrast threshold for deblending of 0.005 demerges (in all three passbands) CFg into two detections, with the flux apportioned between CFgA and CFgB. Table 1 gives Kron magnitudes for the deblended CFgA and CFgB. In F410M, the centroids are  separated by 0.335 arcsec, with CFgB 
0.224 arscec south and 0.249 arcsec east of CFgA. To estimate colours, 
SExtractor is used in the double-image mode to measure the  magnitudes of the deblended components in F410M-fitted Kron apertures. Absolute magnitudes are estimated in Section 5.3.

\begin{table}
\caption{Kron magnitudes for the two deblended components of the CFg, and colours measured in Kron apertures fitted in F410M.}
\begin{tabular}{lccccccc}
\hline
        & CFgA & CFgB \\
\hline
 $m_{410}$ & $23.50\pm 0.08$ & $24.63\pm 0.14$\\
 $V_{555}$ & $24.67\pm 0.04$ & $24.34\pm 0.03$\\ 
 $I_{814}$ & $24.24\pm 0.04$ & $24.08\pm 0.04$\\
$m_{410}-V_{555}$ & $-0.74\pm 0.08$ & $0.05\pm 0.14$ \\
$V_{555}-I_{814}$ & $0.43\pm 0.01$ & $0.22\pm 0.01$
 \\
\hline
\end{tabular}
\end{table}

The surface brightness profiles of CFgA and CFgB are investigated individually, using `isophote', by centering the isophotes on the centroid position of one component
and masking out  the area covered by the other (by eye, dividing approximately at the saddle-point). This analysis is repeated in all three passbands, fitting a new set of isophotes in each.  Figure 8 shows the resulting profiles, with the best-fitting exponential and bulge models, and
Table 2 gives the parameters of the best-fit profiles.

(The $I$ band images are of slightly lower resolution, presumably as the number of exposures is too few for  drizzling to effectively
sample the PSF. This appears to cause a flattening of the $I$-band profiles at
$r<0.08$ arcsec, so the two isophotes within this radius are excluded from the
$I$-band model fits).

\begin{table}
\caption{Best-fitting exponential profiles (central surface brightness
$\mu_0$, scalelength ($r_{exp}$) and bulge-profiles (central surface brightness
$\mu_0^b$, effective radius ($r_e$) for CFgA and CFgB, as plotted on Figure 8, with the rms residuals ($\sigma_{res}$) of each fit in magnitudes.}
\begin{tabular}{lccccccc}
\hline
Galaxy & Band & \multispan{3} \hfil Exponential \hfil & \multispan{3} \hfil Bulge \hfil  \\
     &     &  $\mu_0$ & $r_{exp}$ & $\sigma_{res}$ & $\mu_0^b$ & $r_{e}$
& $\sigma_{res}$ \\
\hline
CFgA & $m_{410}$ & 21.67 & 0.23 & 0.11 & 21.55 & 19.99 & 0.24 \\
CFgA & $V_{555}$ & 22.18 & 0.20 & 0.12 & 19.63 & 5.00 & 0.30 \\ 
CFgA & $I_{814}$ & 21.61 & 0.20 & 0.17 & 18.97 & 4.69 & 0.37 \\
CFgB & $m_{410}$ & 22.06 & 0.21 & 0.11 & 20.20 & 14.1 & 0.18 \\
CFgB & $V_{555}$ & 21.53 & 0.13 & 0.41 & 17.59 & 0.84 & 0.17 \\
CFgB & $I_{814}$ & 21.56 & 0.15 & 0.21 & 17.88 & 1.24 & 0.09 \\
\hline
\end{tabular}
\end{table}

These profiles show that:

(i) On the basis of its UV continuum (presumably from young, massive stars) morphology, CFgA is an exponential galaxy (i.e. most likely a disk) with consistent scalelengths in F555W and $I$, and no evidence for a significant bulge component or central point-source. The F555W $r_{exp}$ corresponds to 
2.42 $h_{50}^{-1}$ kpc.
 
 (ii) The F410M flux from CFgA follows an exponential profile of similar scalelength, indicating that the $\rm Ly\alpha$ emission closely traces the distribution of 
young stars.

(iii) In contrast, the broad-band profiles of  CFgB are closer to a bulge than an exponential, showing an obvious concavity not seen for CFgA. The $V_{555}$ bulge-model $r_{e}$ corresponds to 10.2 $h_{50}^{-1}$ kpc, although this will be an overestimate if there is also an exponential disk component.
 
(iv) The F410M profile of CFgB appears closer to an exponential. Furthermore, CFgB appears smaller in F410M. One possibility is that CFgB is primarily a bulge galaxy with a relatively small
disk component, but only the disk is a $\rm Ly\alpha$ source.

\section{The QSO PHL 957 in F410M}

The QSO PHL 957, at $z=2.681$ is detected close to the RA $01^h03^m11.28^s$ Dec. $+13:16:16.95$ position (equinox 2000.0) of Bunker et al. (1995), originally from Hewitt and Burbridge (1987). It is 
the brightest source on the F410M image, with an aperture magnitude  $m_{410}=17.86\pm 0.01$.  It is saturated on our broad-band images, but through the F410M filter its peak intensity is only $\sim 7$ per cent of the chip saturation level, allowing its profile to be studied.

In F410M, the QSO, with a Gaussian FWHM of $\sim 0.18$ arcsec, is clearly point-source dominated, but it is possible that subtraction of a normalized PSF may reveal the underlying host galaxy, if it is sufficiently luminous
at $\lambda_{rest}\simeq 1110\rm \AA$. Secondly, if the $\rm Ly\alpha$ absorbing cloud on the line-of-sight to the QSO contains a galaxy, it may be revealed close to the QSO position.  

To investigate these possibilities, a model F410M WFPC2 PSF from `Tinytim' is resampled (drizzled) into the smaller pixels of our stacked data,  registered (by cross-correlation) to the QSO position, and normalized to the same flux as the QSO within a 1.0 arcsec radius aperture. We then (i) subtract the normalized PSF from the image of the QSO and look for
evidence of any extended structures in the residual, (ii) compare the QSO and PSF radial profiles, using `isophote'.
\begin{figure}
\psfig{file=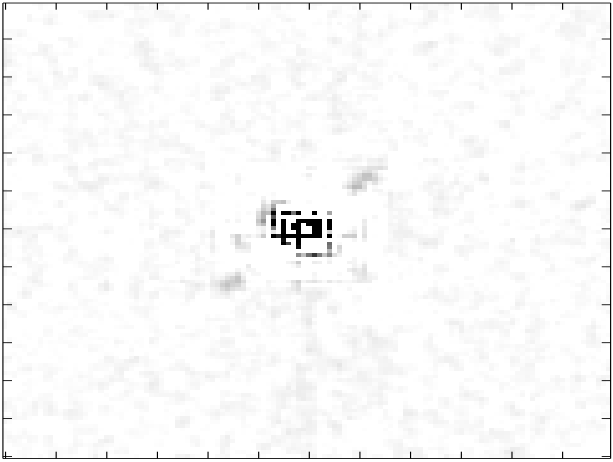,angle=90,width=95mm}
\caption{Greyscale plot of the $6.3\times 6.3$ arcsec area centred on the QSO
 PHL 957 in F410M, after subtraction of a PSF normalized to the QSO intensity.  No significant emission residuals are detected.}
\end{figure}

Figure 9 shows a greyscale plot of the PSF-subtracted QSO image. There is an irregular ring pattern of positive and negative regions near the centre, but this may be due simply to slight differences beweeen the model PSF and that of our data (which will inevitably become less sharp  in the process of combining exposures). There are also patches at the positions of the diffraction spikes (which appear to be slightly 
brighter in the data than in the model PSF), but no obvious indications of any real galaxies. 

Figure 10 shows the radial profiles of the QSO and model PSF. The observed QSO profile is slightly flatter than the model at $r<0.12$ arcsec but otherwise very close to it at all radii, and both profiles have half-light radii of 
$\simeq 3$ pixels $\simeq 0.125$ arcsec.
Again, the QSO appears consistent with a pure point-source.
However, the pixel-to-pixel noise in the background ($\sigma$) is greatly increased near the QSO centre. Figure 10 also shows the pixel-to-pixel rms intensity variation,  $\sigma$, as a function of distance from the AGN, and for comparison, the surface brightness profile of CFgA. 
 To be visible at $>2\sigma$ against the PSF of the AGN, a host galaxy would  require a surface brightness 1.5--2.0
magnitudes brighter than CFgA. Similarly, we estimate by eye that an image of the CFg, superimposed on the PSF-subtracted QSO,  had to be increased in normalization by a factor of $\sim 5$ before there is any visible sign of a galaxy.
\begin{figure}
\psfig{file=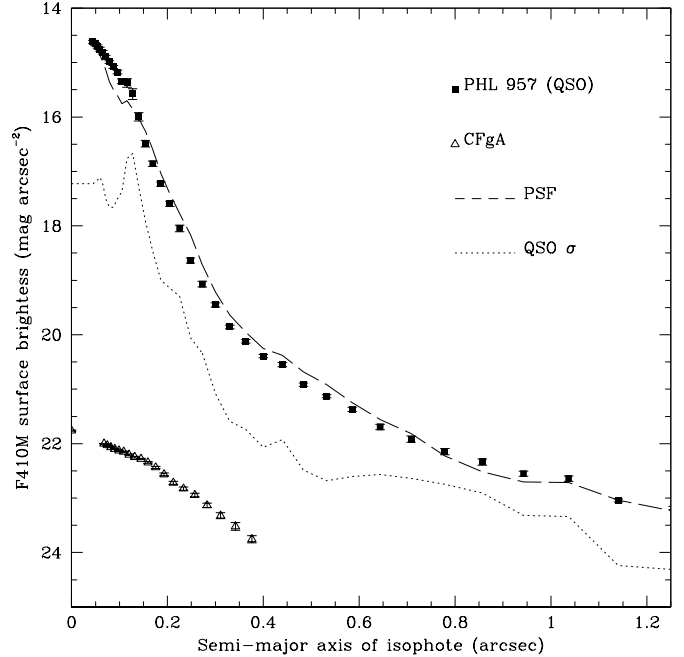,width=95mm}
\caption{F410M surface brightness as a function of semi-major axis, for the QSO
PHL 957 (solid squares), the `TinyTim' model PSF for F410M (long-dashed) and 
the CFg (open triangles). The dotted line shows the rms pixel-to-pixel intensity variation on the observed QSO profile.} 
\end{figure}

We conclude that PHL 957 appears on our F410M image to be consistent with a pure point-source. However, a QSO host galaxy would need to be at least $\sim 2$ mag brighter than CFgA at $r\sim 0.2$--0.4 arcsec to be visible.  At this upper limit, the $\lambda\simeq 1100\rm \AA$ luminosity of the host galaxy would still be exceeded by a factor $\sim 20$ by the AGN. The host galaxies of radio-quiet QSOs are typically $\sim 1$ mag above $L^*$ (McLure et al. 1999), so the 
non-detection may not be unexpected. 

We also find no visible emission from the intervening DLA. A few DLAs have identified optical counterparts, typically $\sim 8$--16$h^{-1}_{50}$ kpc from the QSO (M\o ller and Warren 1993; Fynbo et al. 1999), $\sim 0.65$--1.3 arcsec at the CFg redshift, although these examples differ from the DLA under consideration in that their absorption is at the same redshift as the QSO.
Any galaxy of similar surface brightness to the CFgA would become visible if centred $>0.45$ arcsec from the AGN. If offset by 0.65--1.3 arcsec, it would have to be fainter than CFgA to avoid detection, by as much as $\sim 2 $ mag towards the upper limit of this range. Any galaxy at the centre of the DLA must therefore be compact and  almost concentric with the QSO, or of a relatively low surface brightness in the rest-frame UV (see Section 7.3). 

\section{Galaxy Colours and ${\rm Ly}\alpha$ Emission}
\subsection{Expected Equivalent Widths}
Star-forming galaxies produce $\rm Ly\alpha $ emission primarily from circumstellar HII regions, excited by Lyman continuum ($\lambda<912\rm \AA$) UV from the hottest ($T_{eff}>30000$K) and most massive stars. In the absence of absorption or scattering, the fluxes in both $\rm Ly\alpha$ and the underlying continuum approximately follow the
immediate star-formation rate, and hence the equivalent width $W(\rm Ly\alpha)$ would be relatively insensitive to star-formation history. Charlot and Fall (1993) predict $W(\rm Ly\alpha)$ to decrease during the first $\sim 30$ Myr of star-formation, but thereafter remain approximately  constant in a steadily star-forming galaxy.   

The Bruzual and Charlot `bc96' models (e.g. Charlot, Worthey and Bressan 1996), which assume a Salpeter IMF and 
 neglect the effect of dust, predict $W(\rm Ly\alpha)\simeq 100\AA$
for all steadily star-forming galaxies of age $>0.1\rm Gyr$, increasing to
$W(\rm Ly\alpha)\simeq 150$--$\rm 300\AA$ only in the initial $\sim 12$ Myr.
The `Pegase' models of Fioc and Rocca-Volmerange (1997) incorporate observationally-based dust extinction and a steeper IMF at $M>6M_{\odot}$, and therefore predict a lower $W(\rm Ly\alpha)$. They assume that 70 per cent of Lyman continuum photons are absorbed by the nebulae, and thus may be re-radiated as $\rm Ly\alpha$. Their template spectra give  $W(\rm Ly\alpha)\simeq 32\AA$  -- where $W(\rm Ly\alpha)$ is defined relative to the continuum at $1165<\lambda<1265\rm \AA$  -- for steadily star-forming galaxies of age $>0.1$ Gyr. As in the `bc96' models, $W(\rm Ly\alpha)$ 
increases by factors 2--4 in the very early ($<12$ Myr) stages of star-formation.
Lastly, the low-metallicity Salpeter IMF models of Kudritzki et al. (2000) predict, with 70 per cent Lyman continuum absorption, upper limits (i.e. with no extinction of  $\rm Ly\alpha$) of $W(\rm Ly\alpha)\simeq 160\AA$ for a 3 Myr 
starburst and $W(\rm Ly\alpha)\simeq 75\AA$ for continuous star-formation.
  
Observationally, some local star-forming galaxies  with  $\rm [O/H]<-0.8$ solar show emission of $W(\rm Ly\alpha)=30$--$120\rm \AA$ 
 (Hartmann et al. 1988; Charlot and Fall 1993; Giavalisco et al. 1996). Significantly, the DLA the CFg is associated with is below this abundance threshold (Wolfe et al. 1994). In more metal-rich galaxies, $W(\rm Ly\alpha)<30\rm \AA$ 
and may be decreased to zero.  The situation at high redshifts may be similar;
 Steidel et al. (1999) estimate  only 20--25 per cent of  star-forming Lyman break galaxies at $z\simeq 3.09$ have  $W(\rm Ly\alpha)> 20\rm \AA$, and Lowenthal et al. (1997) detect $\rm Ly\alpha$ emission in only 6/13 $z\sim 3$ galaxies, with these giving a relatively moderate  $W(\rm Ly\alpha)=6.2$--$\rm 34.1\rm \AA$.
It appears that for real star-forming galaxies, the moderate $W(\rm Ly\alpha)$ of the Pegase models may be typical, but there is a very wide range in $\rm Ly\alpha$ properties, perhaps with the `bc96' models approximating an upper limit.

\subsection{Model Galaxy Colours}
Fioc and Rocca-Volmerange (1997) present a set of template spectral energy distributions (continuum and emission-line fluxes) produced using their Pegase models, to fit the observed spectral properties of 8 types of present-day galaxy. Model galaxy SEDs are given at 68 time steps of evolution.  
Using the `Pegase' template SEDs and the response functions of our 3 filters,
we model the observer-frame $m_{410}-m_{555}$ and  $m_{F555}-I$ colours of the evolving galaxies. For the assumed cosmology, the redshift of observation ($z=2.3128$) of the CFg corresponds to a lookback time 13.226 Gyr. 
\begin{figure}
\psfig{file=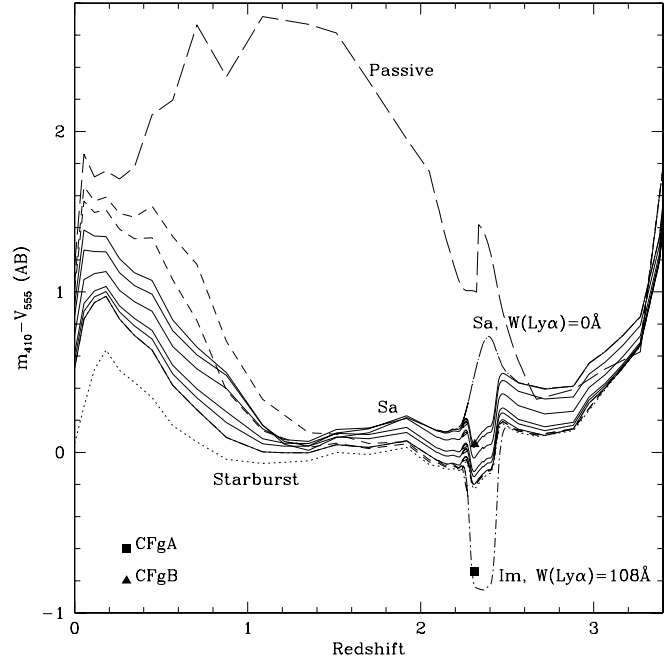,width=95mm}
\caption{Observed $m_{410}-V_{555}$ as a function of redshift for Pegase 
models of passive galaxies (long-dash), E and S0 galaxies (short-dash, upper and lower), Sa/Sb/Sc/Sd and Im spirals (solid, top to bottom), and a  1 Gyr age starburst (dotted) galaxy. Also shown are the Sa model with $W(\rm Ly\alpha)$
reduced to zero (dot-short dash) and the Im model with  $W(\rm Ly\alpha)$
increased to $108\rm \AA$ (dot-long dash).}
\end{figure}  

In the reddest model (`Passive')
considered here, galaxies begin 
star-formation 16 Gyr ago, at $z=6.36$, and cease 
1 Gyr later at $z=4.04$, thereafter evolving passively. In other models the SFR decreases in an approximately exponential fashion with timescales lengthening from E to Im. The E and S0 models form 16 Gyr ago, the Sa, Sb, Sbc, Sd and Im models 15 Gyr ago. We also compute a Burst model in which the SED of a 1.0 Gyr age constant-SFR  burst is redshifted without evolving, to represent the bluest locally observed galaxies (e.g. Gronwall and Koo 1995).

The F410M flux is computed in two separate parts, (i) the $\rm Ly\alpha$ contribution from the product of the Pegase model $\rm Ly\alpha$ flux and the  filter response function at its redshifted wavelength, and (ii) the integration of the continuum over F410M.
Figure 11 shows the predicted $m_{410}-V_{555}$ as a function of redshift.
At $z\sim 2.3$--2.4, all the models (except the $W(\rm Ly\alpha)=0$ Passive model) have $W(\rm Ly\alpha)\simeq 32\rm \AA$, which produces  a blueward shift  in $m_{410}-V_{555}$. An Sa model with $W(\rm Ly\alpha)$ set to zero (shown on Figure 11) becomes redder by 0.46 mag at the CFg redshift and 0.56 mag at $z=2.38$.
Also included is the contribution of the [OII]$3727\rm \AA$ line, which is redshifted through F410M at $z\sim 0.1$, but its effects are small,  $\Delta(m_{410}-V_{555})=-0.16$ mag
mag for starbursts and $-0.10$ mag for spirals, and star-forming galaxies at this redshift remain redder than those at $z>1$.

None of these models approach the extremely blue colour of CFgA. However, a
low-metallicity, dust-free `bc96' model predicts  $W(\rm Ly\alpha)=108\rm \AA$
for a constant-SFR galaxy at $z\sim 2.3$, and is represented on Figure 11 as the Pegase  
Im (i.e. a near-constant SFR) model with $W(\rm Ly\alpha)$ increased to $108\rm \AA$. This gives $m_{410}-V_{555}=-0.84$ at the redshift of CFgA and hence could account for its colour.  
\begin{figure}
\psfig{file=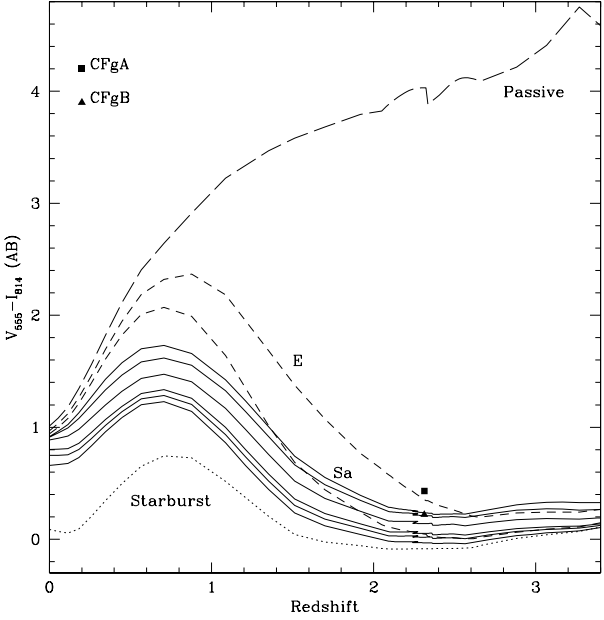,width=89mm}
\caption{Observed $V_{555}-I_{814}$ as a function of redshift for Pegase 
models of passive galaxies (long-dash), E and S0 galaxies (short-dash, upper 
and lower), Sa/Sb/Sc/Sd and Im galaxies (solid, top to bottom), and a  1 Gyr age starburst (dotted).}
\end{figure}  

Figure 11 shows that firstly, at $z\simeq 2.3$--2.4, $m_{410}-V_{555}$ is much more sensitive to $W(\rm Ly\alpha)$ 
than to star-formation history (over the range from Sa to Burst models). Secondly, while starburst galaxies at $0.8<z<2.3$ may be as blue as   $m_{410}-V_{555}\sim -0.05$, any galaxy with $m_{410}-V_{555}< -0.2$ is likely to be a strong $\rm Ly\alpha$ emitter at $2.27<z<2.44$, and  it should be possible to use this colour criterion to select for these (Section 6.1). 

Figure 12 shows the $V_{555}-I_{814}$ colours of the models. At $z\sim 2.3$ the star-forming models lie in the
$V_{555}-I_{814}=-0.1$--0.25 range, but in real galaxies dust extinction may produce larger variations. This colour will give an indication of the star-formation history and/or dust extinction of candidate $\rm Ly\alpha$ sources, and thus help to refine estimates of $W(\rm Ly\alpha)$.

\subsection{Interpreting the CF Galaxy Colours}
We measured  
$m_{410}-V_{555}=-0.74\pm 0.08$ and $V_{555}-I_{814}=0.43\pm 0.01$ for CFgA and 
$m_{410}-V_{555}=0.05\pm 0.14$ and $V_{555}-I_{814}=0.22\pm 0.01$ for CFgB. These are now interpreted by comparison with the Pegase template spectra. 

Assuming CFgA to be a rapidly star-forming disk galaxy, the most appropriate Pegase model is probably a late-type spiral or Im.
The $m_{410}-V_{555}$ is consistent with the Im model with
$W(\rm Ly\alpha)=93\pm 11\rm \AA$, but the $V_{555}-I_{814}$ is much redder than the Im and spiral models. Bunker et al. (1994) detected the $\rm H\alpha$ emission line of the CFg and estimated that the $\rm Ly\alpha/H\alpha$ ratio is reddened by 1.27 ($\pm 0.3$) mag relative to a dust-free starburst. Assuming the  Calzetti et al.(1995) extinction law, this is  $A_V\simeq 0.49$ mag and when added to the Pegase spiral models at $z=2.3$, gives reddening of $\Delta(m_{410}-V_{555})\simeq 0.26$ mag and $\Delta(V_{555}-I_{814})=0.33$ mag.
The `dereddened' colours of CFgA are then $m_{410}-V_{555}\simeq -1.00$ and $E(V_{555}-I_{814})=0.10$, corresponding to the Sc model with 
 $W(\rm Ly\alpha)=151\pm 16\rm \AA$.

For CFgB, the measured colours are consistent with the unmodified Pegase Sa model, with $m_{410}-V_{555}$ best-fit by 
$W(\rm Ly\alpha)=33\pm 13\rm \AA$.Combining these estimates for the two components, weighted by $m_{410}$, gives $W(\rm Ly\alpha)=120\pm 17\rm \AA$ for the whole CFg, consistent with the 
 $W(\rm Ly\alpha)\simeq 140\rm \AA$ estimated (Lowenthal et al. 1991) from the MMT spectrum. 

The observed and model colours can also be used to estimate the $\rm Ly\alpha$ luminosities. If CFgA is represented by the Sc model with 0.26 mag of additional reddening, and CFgB by the Sa model, models with $W(\rm Ly\alpha)$ set to zero predict  $m_{410}-V_{555}=0.60$ and 0.51, respectively,  at $z=2.3128$.
The observed colours indicate that 70.9 per cent and 34.5 per cent of their respective F410M fluxes are $\rm Ly\alpha$, rather than continuum. The $\rm Ly\alpha$ components of their 
F410M magnitudes are then $m_{410}=23.87\pm 0.08$ for CFgA and $25.78\pm 0.14$
for CFgB, equivalent to 
$10.3\times 10^{-30}$ and $1.77\times 10^{-30}$ ergs $\rm s^{-1}cm^{-2}Hz^{-1}$  averaged over the bandwidth.

 The restframe photometric bandwidth in   frequency units is  $c(1+z)\Delta\lambda/\lambda_{pivot}^2$ -- where for F410M,  $\Delta\lambda=93.73\rm\AA$ and  $\lambda_{pivot}=4092.7\rm\AA$ (from image header) -- giving $5.558\times 10^{13}$ Hz. Multiplying, the $\rm Ly\alpha$ line fluxes are then $5.71\pm 0.43\times 10^{-16}$ ergs $\rm s^{-1} cm^{-2}$ for CFgA and $0.984\pm 0.127\times 10^{-16}$ ergs $\rm s^{-1}
cm^{-2}$ for CFgB.
 This is consistent with the Lowenthal et al. (1991) estimates of 
$6.5\times 10^{-16}$ (Fabry-Perot) and $5.6\times 10^{-16}$ (MMT spectrograph) ergs $\rm s^{-1} cm^{-2}$ for the whole source. For $q_0=0.05$, the $\rm Ly\alpha$ luminosities are
$5.21\times 10^{43}h_{50}^{-2}$ ergs $\rm s^{-1}$ for CFgA and $8.99\times 10^{42}$ ergs $\rm s^{-1}h_{50}^{-2}$ for CFgB (this is discussed further in Section 7.1).
 
Lastly, the Pegase models can be used to estimate absolute magnitudes, e.g.
$M_B$ in the rest-frame blue-band. At $z=2.3128$ the Sa model spectrum gives a k-correction $B_{rest}-I_{obs}=-0.05$ mag and the reddened Sc model, $B_{rest}-I_{obs}=-0.14$ mag. Adding the demerged Kron magnitudes of $I_{814}=24.24\pm 0.04$ for CFgA and $I_{814}=24.04\pm 0.04$ for CFgB,  the distance modulus $-47.20+5~{\rm log}~h_{50}$ and these corrections, we estimate $M_B=-23.10+5~{\rm log}~h_{50}$ for CFgA and $-23.21+5~{\rm log}~h_{50}$ for CFgB. Both components are approximately 2 magnitudes brighter than the present-day
$L^*$. 

\section{Other $\rm Ly\alpha$ Emitting Galaxies}
One aim of imaging the CFg field in narrow band and continuum filters was to identify other $\rm Ly\alpha$ emitters at the same redshift, and thus determine if the CFg is in a cluster of these sources. Pascarelle et al. (1998) found evidence of a concentration of $\rm Ly\alpha$ emitters near a high-redshift radio galaxy, and we would expect a local overdensity to be associated with a damped $\rm Ly\alpha$ cloud.
 
\subsection{Selection}
The models of Section 5.2 predicted that high $W(\rm Ly\alpha$) galaxies at $z\sim 2.3$--2.4 would be identifiable by a colour
$m_{410}-V_{555}<-0.2$. 
A total of 71 sources are detected on the three chips of the F410M image. As before, colours are measured from Kron magnitudes in apertures matched to the F410M detection. There are a  total of 15, including the CFg, with $m_{410}=V_{555}<-0.2$, but the brightest two are PHL 957 and a bright star
-- both saturated in F555W giving an unreliable colour -- and three of the faintest appeared to be spurious detections. Excluding these leaves the CFg and 9 fainter sources.

\begin{figure}
\psfig{file=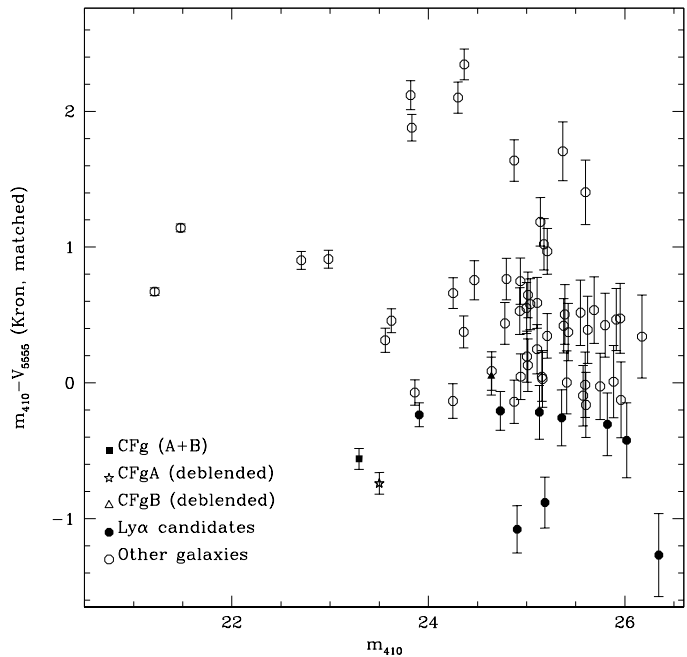,width=95mm}
\caption{Observed $m_{410}-V_{555}$ for all F410M detections (from Kron magnitudes with detection in F410M), against $m_{410}$ magnitude. Filled symbols indicate the candidate $\rm Ly\alpha$ emitters. The CFg is plotted twice, as a single source and as the  deblended galaxies.}
\end{figure}  

  Table 3 gives the F410M magnitudes and colours  of the candidate $\rm Ly\alpha$ emitters, numbered as `Lc:n' where c is the chip number and n the SExtractor detection number, and also the $I$ magnitudes in $I$-band fitted Kron apertures, which (if the sources are at $z\sim 2.3$) indicate $\lambda \sim 2500\rm \AA$ luminosites ($\sim$ mass of young stars) independent of the $\rm Ly\alpha$ properties. Figure 13 shows $m_{410}-V_{555}$ against $m_{410}$ for the candidate $\rm Ly\alpha$ emitters and the other F410M detections.
\begin{table}
\caption{The F410M magnitudes and colours (Kron, in F410M apertures), and $I_{814}$
magnitudes (Kron, measured in $I$ apertures) of 
candidate  $\rm Ly\alpha$ emitters at
$z\sim 2.3$--2.4 ($nd=$ not detected in $I$), with CFgA/B for comparison.} 
\begin{tabular}{lcccc}
\hline
\smallskip
Source No. & $m_{410}$ & $m_{410}-V_{555}$ & $V_{555}-I_{814}$ & $I_{814}$ \\
L2:03 & 25.13 & $-0.22\pm 0.20$ & 0.64 & 24.39 \\
L2:09 & 25.19 & $-0.88\pm 0.19$ & 0.12 & 25.99 \\
L2:12 & 24.73 & $-0.21\pm 0.14$ & 0.83 & 23.25\\
L2:16 & 26.35 & $-1.27\pm 0.31$ & -1.09 & nd \\
L3:10 & 23.91 & $-0.24\pm 0.09$ & 0.06 & 24.02\\
L3:19 & 25.36 & $-0.26\pm 0.21$ & 0.60 & 24.74\\
L4:07 & 25.82 & $-0.31\pm 0.23$ & -0.15 & 25.80 \\
L4:11 & 24.90 & $-1.08\pm 0.17$ & -0.04 & 25.74 \\
L4:12 & 26.02 & $-0.42\pm 0.28$ & 1.28 & 24.78 \\
CFgA  & 23.50 & $-0.74\pm 0.08$ & 0.43 & 24.24 \\
CFgB  & 24.64 & $0.05\pm 0.14$ &  0.22 & 24.08 \\
\hline
\end{tabular}
\end{table}

Source L2:12, which only just satisfies the colour criterion, is uncertain -- the F410M detection forms only part of a much larger, morphologically complex galaxy, the remainder of which is redder in $m_{410}-V_{555}$, and the large $I$-band FWHM of 1.68 arcsec may be implausible for an object at this redshift. It may be a lower redshift starburst galaxy. 
The other 8 candidate sources have continuum sizes and magnitudes consistent with $z\sim 2.3$ galaxies, and are definitely blue in $m_{410}-V_{555}$.

\subsection{Colours and Morphology}

CFgA is $>0.4$ mag brighter in F410M than any of the candidate sources, and remains the most powerful $\rm Ly\alpha$ emitter in the field and redshift range of observation. The $I_{814}$ magnitudes of the candidate sources (excluding L2:12) range from 0.22 mag brighter to $>2$ mag fainter than CFgA; assuming a similar k-correction as for CFgA gives their mean absolute magnitude as
 $M_B\sim -22.3$, $\sim 1$ mag brighter than  the present-day $L^*$.

Figure 14 shows a colour-colour plot ($m_{410}-V_{555}$ against $V_{555}-I_{814}$), for
the F410M detections. Of the candidate $\rm Ly\alpha$ emitters, L3:10 and L4:07 lie only slightly below the Pegase models at $z=2.3128$, suggesting $W(\rm Ly\alpha)\simeq 30$--$50\rm \AA$. 
Four sources (including L2:12) have $V_{555}-I_{814}>0.5$, and if genuinely at $z\sim 2.3$ are probably  dust-reddened and hence stronger $\rm Ly\alpha$ emitters than would be thought from their  $m_{410}-V_{555}$ alone. 
\begin{figure}
\psfig{file=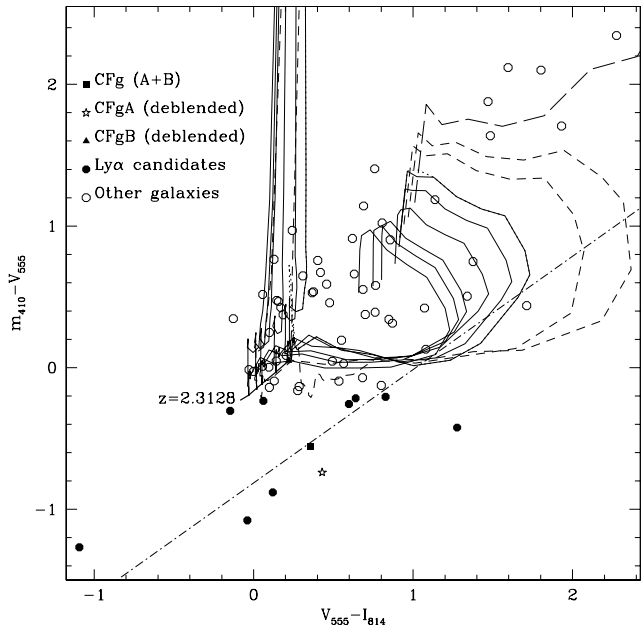,width=91mm}
\caption{Colour-colour plot of all F410M detections, with the CFg plotted as a single source and as two deblended galaxies. The long-dashed, short-dashed and solid lines represent the Pegase models as on Figures 11 and 12.  The model galaxies follow these loci clockwise with increasing redshift, from $z=0$ to 
$z\sim 3.5$. The heavy solid diagonal line connects the positions of the Pegase spiral models at $z=2.3128$, from Im (left) to Sa (right). The dot-dash diagonal line intersects an Im model with $W(\rm Ly\alpha)=108\AA$ and follows the slope of the dust-reddening vector.} 
\end{figure}  

 The dot-dash line passes through the colours of an 
Im model with $W(\rm Ly\alpha)=108\rm \AA$ (the `bc96' prediction), and follows the slope of the dust-reddening vector,  estimated as $\Delta(m_{410}-V_{555})\simeq 0.80 \Delta(V_{555}-I_{814})$. Several candidate $\rm Ly\alpha$ emitters lie close to this line - one possible interpretation is that they (and the CFg) have rather similar
 $\rm Ly\alpha$ properties but dust reddening ranging from near-zero to $\Delta(V_{555}-I_{814})\sim 1$ mag. 

\begin{figure}
\psfig{file=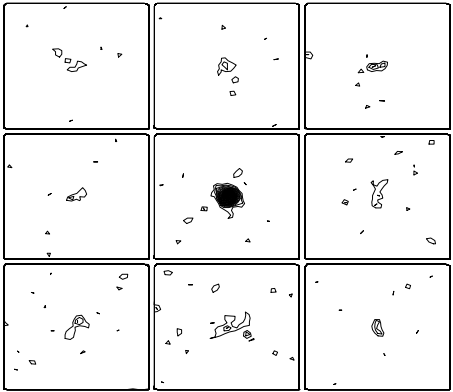,width=95mm}
\caption{Contour plots of $2.1\times 2.1$ arcsec areas of the F410M image, centred on each of the 9 $\rm Ly\alpha$ candidates; (top, left to right) 
L2:03, L2:09, L2:12;(middle) L2:16, L3:10, L3:19; (bottom) L4:07, L4:11, L4:12).
Contours are linearly spaced, the lowest being $2\sigma$ above the sky.}
\end{figure}
 
\begin{figure}
\psfig{file=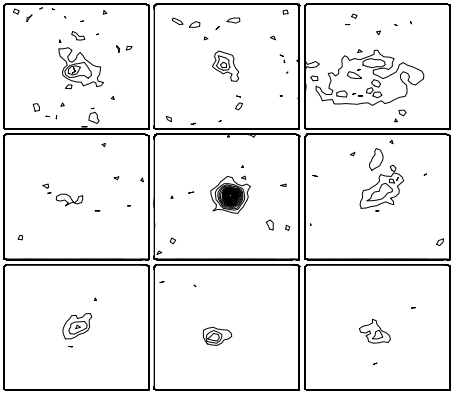,width=95mm}
\caption{As Figure 15 in F555W.}
\end{figure}

Figures 15 and 16 show contour plots of the 9 candidate $\rm Ly\alpha$ sources in F410M and F555W. Most have the typical appearence of Lyman-break objects or `chain galaxies' -- small bright nuclei and a number of secondary intensity peaks, presumably knots of star-formation, within
an asymmetric outer envelope. They are generally more irregular than either component of the CFg, and appear to be single galaxies rather than interacting
pairs. They appear even more `knotty' in F410M than in the broad bands, as expected if the $\rm Ly\alpha$ emission is from multiple star-forming hotspots. However, L3:10, the brightest after CFgA, differs from the others in being a single-peaked, compact, high central surface brightness source in all 3 passbands.   

The F410M sizes and central surface brightnesses of these sources are estimated by fitting exponential profiles to the F410M isophotes, determined using  {\sevensize IRAF} `isophote'.   
  Table 4 gives the best-fit parameters, with our previous results for CFgA/B, and also gives the centroid positions as RA and Dec from WFPC2 astrometry, which is accurate to $<2$ arcsec for the absolute positions and will be much more precise than this for their relative positions. To aid in finding 
the objects, we also give the position of the QSO PHL 0957, in the same 
co-ordinate system.

Some objects, with multiple peaks, show large residuals from an exponential
profile, but none appear to
be bulge galaxies. The profile of L3:10  resembles the PSF, and subtracting a normalized PSF from its image appeared to leave no significant residuals. Hence L3:10 must be almost a pure point-source, and   may be a QSO. The $r_{exp}$ of the other candidate sources are generally comparable to the CFgA. The brighter F410M magnitude of the CFgA is not the result of a larger size, but of a higher surface brightness, $>1$ mag above most of these galaxies. 
 
  \begin{table}
\caption{The RA and Dec positions (equinox 2000.0) from WFPC2 astronomy, best-fitting exponential
scalelengths ($r_{exp}$, in arcsec) F410M central surface brightnesses ($\mu_{410}$, in AB mag $\rm arcsec^{-1}$), of candidate $z\sim 2.3$--2.4 $\rm Ly\alpha$ emitters, with CFgA/B and PHL 957 for comparison. No corrections are applied for the PSF, which for a pure point-source gives $r_{exp}=0.063$ arcsec.} 
\begin{tabular}{lcccc}
\hline
\smallskip
Source  & R.A. & Dec. & $r_{exp}$ & $\mu_{410}$ \\
L2:03 & $01^h 03^m 09.87^s$ & $13:16:38.84$ & 0.39 & 23.38 \\
L2:09 & $01^h 03^m 09.88^s$ & $13:16:29.97$ & 0.31 & 22.86 \\
L2:12 & $01^h 03^m 10.13^s$ & $13:16:19.77$ & 0.21 & 22.57 \\
L2:16 & $01^h 03^m 12.86^s$ & $13:15:45.31$ & 0.09 & 22.38 \\
L3:10 & $01^h 03^m 07.67^s$ & $13:16:30.61$ & 0.07 & 19.92 \\
L3:19 & $01^h 03^m 05.55^s$ & $13:16:40.85$ & 0.35 & 22.99 \\
L4:07 & $01^h 03^m 09.15^s$ & $13:17:17.24$ & 0.17 & 22.71 \\
L4:11 & $01^h 03^m 10.17^s$ & $13:18:00.53$ & 0.39 & 22.79  \\ 
L4:12 & $01^h 03^m 06.88^s$ & $13:17:44.18$ & 0.23 & 22.93 \\
CFgA  & $01^h 03^m 08.45^s$ & $13:16:41.50$ & 0.23 & 21.67 \\
CFgB  & $01^h 03^m 08.47^s$ & $13:16:41.27$ & 0.21 & 22.06 \\
PHL957 & $01^h 03^m 11.30^s$ & $13:16:17.99$ & 0.063   & 14.62 \\
\hline
\end{tabular}
\end{table}

Figure 17 shows the distribution of the candidate $\rm Ly\alpha$ sources on the field. They appear to be evenly distributed, without obvious clustering about CFg or any other position, and none are closely paired (except the two components of CFg).

\begin{figure}
\psfig{file=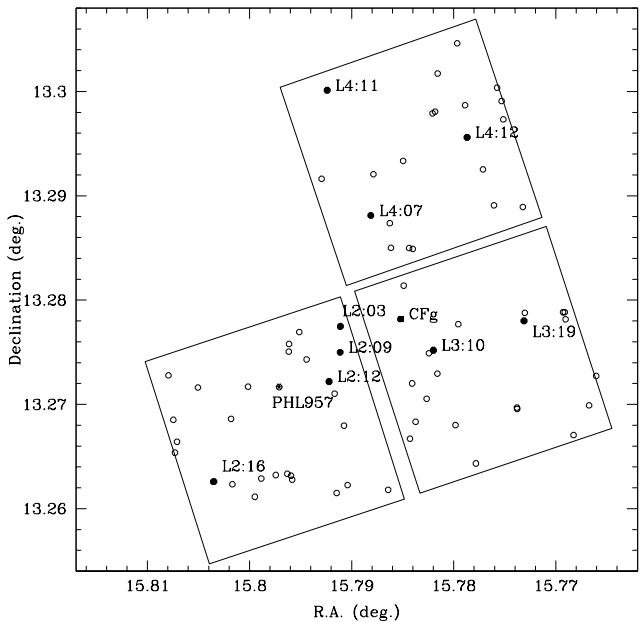,width=90mm}
\caption{Distribution on the sky of the $\rm Ly/alpha$ candidates (solid circles) and other F410M detections (open circles), with the CFg (solid square) and the QSO PHL 957 (asterisk). The three boxes show the observed areas (chips
2, 3 and 4).}
\end{figure}
\section{Discussion}
\subsection{The CFg - a Starbursting Merger?}
We have investigated a $z=2.3128$ $\rm Ly\alpha$ emitting galaxy with high-resolution, deep WFPC2 imaging in narrowband $\rm Ly\alpha$ (F410M) and
the UV continuum (F555W and $I$). The galaxy, in all 3 passbands, is resolved into two components of similar size, CFgA and CFgB, with projected separation 0.335 arcsec ($\sim 4.1 h_{50}^{-1}$ kpc). It is probable that the CFg is a merging pair of 
galaxies.  Firstly, both components are extended and moderately elongated; neither resembles a point source or jet that might just be a region of star formation in a single galaxy. Each component,  investigated individually, shows the radial profile of a normal galaxy, CFgA being exponential while CFgB is closer to a bulge. Secondly, the two components  differ in colour, CFgA being much bluer in $m_{410}-V_{555}$ but
slightly redder in $V_{555}-I$, while the colours within each component are relatively uniform.

From the narrowband magnitudes and colours of the deblended components, $m_{410}=23.50\pm 0.08$ and $m_{410}-V_{555}=-0.74\pm 0.08$ for CFgA, $m_{410}=24.64\pm 0.14$ and  $m_{410}-V{555}=0.05\pm 0.14$ for CFgB, we estimate $\rm Ly\alpha$ luminosities of 
 $5.21\times 10^{43}h^{-2}_{50}$ ergs $\rm s^{-1}$ for CFgA and $8.99\times 10^{42}h^{-2}_{50}$ ergs $\rm s^{-1}$ for CFgB.

Kennicutt (1983) found $L(\rm H\alpha)\sim 1.12\times 10^{41} \times$ SFR ($\rm M_{\odot} yr^{-1})$ ergs $\rm s^{-1}$ for spiral galaxies, where SFR is the total star-formation rate for a Salpeter IMF. For star-formation in the absence of dust, $L({\rm Ly}\alpha)\simeq 8.3 L({\rm H}\alpha)$ (Osterbrock 1989, Bunker et al. 1995), so that the SFR would be $L(\rm Ly\alpha)/(9.296\times 10^{41}$ ergs $\rm s^{-1})$, giving 56.0 $h^{-2}_{50}\rm M_{\odot}yr^{-1}$ for CFgA and 9.7 $h^{-2}_{50}\rm M_{\odot}yr^{-1}$ for CFgB. Bunker et al. (1995) estimate that the $\rm Ly\alpha/ H\alpha$ ratio of the CFg is dust-reddened by 1.27 mag, which corresponds to 1.56 mag of $\rm Ly\alpha$ extinction and increases the estimated SFR of CFgA to $\sim 236 h^{-2}_{50} \rm M_{\odot} yr^{-1}$. 

The  SFR in CFgA is very high, as might be expected for a gas-rich disk galaxy at the peak of a merger-induced starburst (e.g. Mihos and Hernquist 1996; Read and Ponman 1998).  In contrast, the $L(\rm H\alpha)$ of CFgB is consistent with a SFR in the range of normal spirals (Kennicutt 1983). One possible explanation is that the starbursts in the two galaxies peak at different times due to the bulge morphology of CFgB -- in simulations (e.g. Mihos and Hernquist 1996; Tissera 1999), the presence of a massive bulge  delays the SFR maximum to a later stage of a merger.

\subsection{Surface Brightness Evolution}
We fitted radial profiles to both components in all three passbands, and 
estimated rest-frame blue-band absolute magnitudes from the broad-band magnitudes and Pegase model spectra. These two measurements provide an estimate of the surface brightness evolution, independent of the $\rm Ly\alpha$ properties, and of $q_0$ and $H_0$.
 The size-luminosity relations for local 
galaxies, from Roche et al. (1998) with $B_{Vega}=B_{AB}+0.077$, are
$${\rm log}~r_{hl}=-0.2M_B-3.235$$ for spirals (derived from Freeman 1970) and $${\rm log}~r_{e}=-0.3M_B-5.594$$ for ellipticals (from Binggeli, Sandage and
Tarenghi 1984). 

For CFgA, best-fitted by a pure exponential profile, we estimate $r_{exp}=2.42
h_{50}^{-1}$ 
and $M_B=-23.10 +5~{\rm log}~h_{50}^{-1}$.
For an exponential profile $r_{hl}=1.68 r_{exp}= 4.07h_{50}^{-1}$ kpc for CFgA, and for a local spiral of this size $M_B=-19.22+5 {\rm log}~h_{50}$. 
This observed luminosity is greater by 3.88 mag or a factor 35.6.
This surface brightness evolution exceeds the evolutionary brightening of the Pegase  spiral models , e.g.  $\Delta(M_B)=-0.75$ mag for Sc and $-1.80$ mag for Sa galaxies from $z=2.3128$ to
 $z=0$, but is consistent with  a 1 Gyr starburst ending at $z\sim 2.3$, which  would subsequently fade by 3.98 mag.
Hence the continuum surface brightness of CFgA, like the $\rm Ly\alpha$ luminosity, implies that the SFR greatly exceeds the time-averaged rate and that an intense short-term burst is occurring. 

Lyman break galaxies tend to be small in size but very high in surface brightness. Roche et al.(1998) estimated that the rest-frame blue surface brightness of 16 Hubble Deep Field galaxies at $2.26<z<3.43$ averaged $2.79\pm 0.31$ mag higher than $z<0.35$ galaxies, with $\sim 1$ mag dispersion. Lowenthal et al. (1997) found the mean $r_{hl}$ of $z\sim 3$
Lyman break galaxies to be 3.6 kpc with $M_B$ ranging from -21.6 to -23.1. The surface brightness of CFgA is still a factor $\sim 2$
higher than the average for Lyman break galaxies (in these flux-limited samples), and near the maximum of the
observed range. Again, this is most likely the result of a particularly luminous merger-triggered starburst.

The effective radius of CFgB was estimated as $10.2 h_{50}^{-1}$ kpc, from fitting a pure bulge (${\rm exp}~ [r^{-0.25}]$) profile, and its absolute magnitude as $M_B=-23.20+5 {\rm log}~h_{50}$.  For a local elliptical of this size, $M_B=-22.01+5 {\rm log}~h_{50}$, indicating surface brightness evolution of 1.21 mag. This is slightly less than predicted by the Sa model, but is a lower limit as the presence of any disk component would cause 
$r_e$ to be overestimated. The surface brightness of CFgB, like the $\rm Ly\alpha$ luminosity,  suggests a more moderate evolution for this component, essentially consistent with the Pegase models.

\subsection{Environment of the CFg}
Models predicted that $W(\rm Ly\alpha)>30\rm \AA$ sources at $2.27<z< 2.44$ would be separable from other star-forming galaxies by a colour $m_{410}-m_{555}<-0.2$. We used this colour criterion to search for other $\rm Ly\alpha$ emitters at the same redshift as the CFg, and  found nine candidate sources (in addition to the CFg) to the faint limits of the F410M data. One (L2:12) is suspected to be a foreground galaxy. Of the other 8, it is not yet known which, if any, are physically associated with the CFg, and which are field galaxies distributed over 
$2.27<z<2.44$.  To interpret these results in terms of the environment of the CFg, the total number of candidates must be compared with both random and known cluster fields. 

   Pascarelle et al. (1998) imaged four fields with WFPC2 through F410M, F450W ($B$) and F606M ($V$) filters. The use of the same F410M filter means that $\rm Ly\alpha$ sources will be selected in exactly the same redshift range as in our observations. On a field containing the $z=2.4$ radio galaxy
53W002, 17 galaxies were selected as $\rm Ly\alpha$ candidates, and 
spectroscopy confirmed that some of these did form a cluster with 53W002. 
Three randomly selected fields contained 3, 11 and 4   $\rm Ly\alpha$ candidates (it was thought the second of these had by chance contained a cluster). The F410M exposure times of these fields differed and all were shorter than for our image. To a common magnitude limit $B=25$, they give the numbers of candidates per field as 10, 1, 8 and 2, and we would (counting the CFg as one galaxy) find 4. 

On this basis it seems that the CFg environment is likely to  be less rich than that of 53W002. On the other hand, the number of $\rm Ly\alpha$ candidates in our field may still be about twice that expected from random fields. Detecting $\rm Ly\alpha$ emission is not a good estimator of cluster richness as, even if most cluster galaxies are star-forming at these redshifts,  only a small and uncertain fraction will have strong $\rm Ly\alpha$ emission (Steidel et al. 1999). Mannucci et al. (1998) instead imaged the PHL 957 field through a $1.237\rm \mu m$ filter designed to detect redshifted $\rm [OII]3737\rm \AA$,
together with a broadband filter. This revealed two candidate [OII] sources, close to the CFg in both position and redshift, although just outside the area studied here. They would fall on the PC chip of our image, on which no $\rm Ly\alpha$ sources could be seen. Whether or not these galaxies are significant $\rm Ly\alpha$ emitters, the [OII] fluxes estimated by Mannucci et al. (1998) 
correspond to very high SFRs, similar to the CFg.
Taking this and our findings into account it seems probable that the CFg does lie within some sort of high-redshift group or cluster, in which a large amount of star-formation is taking place.

The CFg is also known to be associated with a
DLA cloud. We attempted, unsucessfully, to find the DLA optical counterpart on our F410M image. On the basis of previous observations (e.g. Fynbo et al. 1999; Djorgovski et al. 1995), it is most likely to lie close ($\sim 1$ arcsec) to the QSO PHL 957, but due to the increased noise in this region of the data, the upper limits on its central surface brightness are relatively high -- about equal to CFgA 0.45 arcsec from the QS0 and $\sim 2 $ mag fainter 1.3 arcsec from it. As CFgA is a particularly high surface brightness galaxy, these limits may allow a spiral or irregular galaxy to have formed in the cloud. The optical counterparts of high-redshift DLAs may typically be of low/moderate luminosity, e.g. the
`S1' object discovered by M\o ller and Warren (1998) at $z=2.81$ is fainter than the CFg, with $I_{AB}=25.3$, the $\rm Ly\alpha$ luminosities of DLAs so far detected are also lower than CFgA, and Fynbo et al. (1999) predict the majority of DLAs to  be at $R>28$.  

\subsection{The Lyman $\alpha$ Properties}
\subsubsection{Comparison with other $Ly\alpha$ Galaxies}
We estimate rest-frame $\rm Ly\alpha$
equivalent widths of $W(\rm Ly\alpha)=151\pm 16\rm \AA$ for CFgA and $W(\rm Ly\alpha)=33\pm 13\rm \AA$ for CFgB. 
The $W(\rm Ly\alpha)$ of CFgA is, on the basis of the Charlot and Fall (1993)
models, close to the maximum possible for a non-AGN starburst, but is not
unprecedented. The most obvious comparison is with the `G2' galaxy ($z=3.428$) described by 
Giavalisco at al. (1995), with $W(\rm Ly\alpha)=175\pm 16\AA$, an estimated SFR of 50--100 $M_{\odot}\rm yr^{-1}$, and a luminosity only slightly lower than the CFg,
$M_B=-22.7$. This object differs markedly from CFgA in showing a compact
($r_e=1.3$ kpc) bulge profile, but as in the CFg there is no evidence of a central point source. 

The candidate $\rm Ly\alpha$ sources on our field are also of lower luminosity than CFgA, but in contrast to `G2', tend to be more irregular and of lower surface brightness (with the exception of the possible AGN, L3:10). Their colours suggest a wide $\sim 30$--$150\rm \AA$ range of $W(\rm Ly\alpha)$, with a distribution approximately centred on a locus corresponding to
$W(\rm Ly\alpha)\sim 108\rm \AA$ and dust extinction $0<A_V<1.5$ mag.  
Similarly, the 35 candidate $\rm Ly\alpha$ sources on the 4 fields of
 Pascarelle et al. (1998), at the same redshifts,  have $m_{410}-B$ and $B-V$ consistent
with $W(\rm Ly\alpha)\sim 20$--$180\rm \AA$ and extinction $0<A_V<2$ mag. These vary in size but most are smaller than either component of the CFg (28/35 have $r_{hl}<0.3$ arcsec).

Malkan, Teplitz and McLean (1996) investigate by spectroscopy a $z=2.498$ galaxy
with a similar $W(\rm Ly\alpha)$ and luminosity to `G2', thought to be a 
$\sim 100$ $\rm M_{\odot}\rm yr^{-1}$ starburst with high dust extinction ($A_V\sim 0.9$). Steidel et al. (1999) identify 15 $\rm Ly\alpha$ emitting galaxies at $z\sim 3.09$, with   W($\rm Ly\alpha)_{median}=67\rm \AA$, and a continuum luminosity function the same shape as for other Lyman break galaxies. 
 At even higher redshifts, Dey et al. (1998) measure $W(\rm Ly\alpha)=95\rm\AA$ for a galaxy at $z=5.34$.
Kudritzki et al (2000) identify 9 $\rm Ly\alpha$ emitters at $z\simeq 3.1$ in a narrow-band survey originally aiming to detect planetary nebulae, and confirm the redshifts by spectroscopy. These sources have $W(\rm Ly\alpha)\simeq 33$--$175\rm \AA$ and $\rm Ly\alpha$ luminosities about 8--66 per cent that of CFgA.
Lastly, Taniguchi and Shoiya (2000) propose that the recently discovered large ($\sim 100$ kpc) $\rm Ly\alpha$  nebulae are powered by primordial ellipticals, with starbursts which can  exceed even CFgA in $\rm Ly\alpha$ luminosity. 
 
Together, these results indicate that (i) a significant number of high redshift
star-forming galaxies have $W(\rm Ly\alpha)$ exceeding the $30\rm \AA$ of the Pegase models, and ranging up to an upper limit of
$\sim 150$--$180\rm \AA$, and (ii) these strong $\rm Ly\alpha$ emitters cover a wide range of luminosity, size, age and dust extinction and include both bulge and disk galaxies. The $W(\rm Ly\alpha)$ and luminosity of CFgA are near the upper limit of this distribution. 

\subsubsection{Interpretation}
As many low metallicity and high-redshift star-forming galaxies reach $W(\rm Ly\alpha)\sim 30\AA$ (Hartmann et al. 1988; Lowenthal et al. 1997),
the moderately high SFR typical at these redshifts  may be sufficient to explain the $\rm 
Ly\alpha$ emission from CFgB. The much greater $W(\rm Ly\alpha)$ of CFgA and
the  other sources discussed above cannot be explained by further reducing the dust -- many of these 
have colours indicating they are more reddened than CFgB -- and there must be an additional property of galaxies influencing $\rm Ly\alpha$ emission.

Kunth et al. (1998) propose that the most important factor in determining the escape of $\rm Ly\alpha$ photons from star-formation is not the dust, but the velocity structure of the gas. Using high resolution spectroscopy, they  found 4 of a sample of
8 starburst galaxies, including the very low metallicity IZw18, showed only a broad absorption trough at the $\rm Ly\alpha$ wavelength. In the 4 with $W(\rm Ly\alpha) >0$, the trough is still present but shifted to the blue side of the emission line, producing an asymmetric `P Cygni' line profile.
This indicates a galactic wind, in which the neutral gas is outflowing at some hundreds of km $\rm s^{-1}$ relative to the star-forming central regions. This velocity difference greatly reduces resonant scattering of the $\rm Ly\alpha$ photons, which would otherwise increase the effects of dust extinction and 
obscure the emission line.

Pettini et al. (1998) perform near-infra-red spectroscopy of 5  galaxies at $z\sim 3$, of which 3 show $\rm Ly\alpha$ in emission. They also find evidence of high-velocity ($\sim 500$ km $\rm s^{-1}$)
large-scale outflows, seen as a blueshifting of metal absorption lines and a 
redshifting of $\rm Ly\alpha$ relative to the H$\beta$ and [OII] emission lines of the star-forming nebulae.
  
Tenorio-Tagle et al. (1999) develop a model of an instantaneous $10^6M_{\odot}$  starburst that produces a rapidly expanding bubble inside a galactic disk.
They predict a sequence in which (a) at the start of the burst, $\rm Ly\alpha$ photons are trapped, 
(b) after $\sim 2.5$ Myr the expanding bubble breaks throught the galactic disk, producing a cone of ionization through which most UV photons can escape, (c) at $\sim 4$ Myr, recombination in the expanding shell produces a secondary $\rm Ly\alpha$ emission line, blueshifted with respect to the first, (d) at $>5$ Myr the ionization front becomes trapped behind the recombining shell and the secondary emission becomes an absorption trough, giving a P Cygni profile, (iv) at $>9$ Myr continuing recombination in the galaxy halo again cuts off the $\rm Ly\alpha $ photons, and for the remainder of the
burst, $\rm Ly\alpha$ is seen only as an absorption feature.

For the more massive and prolonged starbursts in the CFg and other high-redshift sources, the period of $\rm Ly\alpha$ emission is likely to exceed that in this model, although it is probably shorter than the $\sim 100$ Myr
predicted for a much more extended ($\sim 100 \rm kpc$) source 
(Taniguchi and Shioya 2000). While evolutionary timescales may vary, the most important prediction of this model is that powerful starbursts would pass through a phase of strong $\rm Ly\alpha$ emission -- perhaps attaining the $\rm W(Ly\alpha)\sim 150\AA$ of unobscured models (Section 5.1)  -- which is relatively brief compared to 
merging and other dynamical timescales. This would explain why strong $\rm Ly\alpha$ emitters are diverse in morphology and have a similar luminosity function to other star-forming galaxies, but form only a small fraction of any
sample of these (Steidel et al. 1999). 
It also means that CFgA and CFgB need be only slightly mismatched in the timing of their peak SFR to explain their large difference in $W(\rm Ly\alpha$).

A further prediction is that in galaxies near the beginning of the $\rm Ly\alpha$ emission phase (the first $\sim 1$ Myr in the model), the $\rm Ly\alpha$ profile shows one or more secondary emission lines -- in massive starbursts,  a `network of shells` produces a `forest of $\rm Ly\alpha$ in emission`  --  while later stages give a P Cygni line profile. The Lowenthal et al. (1991) MMT spectra (Figure 1 and 2) of the CFg may be of insufficient depth and resolution to unambiguously detect either of these features, but the medium-resolution ($\rm FWHM\simeq 2.6\rm \AA$) spectrum does indicate an intrinsic  Gaussian FWHM of $\simeq \rm 8 \AA$ corresponding to high velocity FWHM of $\sim 600$ km $\rm s^{-1}$. The line is sharp-peaked with no obvious asymmetry. 
Fitting a Voigt model suggests that the profile is closer to a  $\rm FWHM\simeq 6\rm \AA$ Lorentzian than a Gaussian. We hypothesize that CFgA is in the earlier starburst stage, with multiple outflowing shells with high relative velocities giving the appearence of a single broadened $\rm Ly\alpha$ emission line, but no  P Cygni profile. To confirm this, more data are needed; see Section 7.5i.

 We find no positive evidence for an AGN contribution  to the $\rm Ly\alpha$ emission. The F410M flux of CFgA (of which $\sim 70$ per cent is from the $\rm Ly\alpha$ line) appeared to trace the exponential profile
of the UV continuum, with no evidence of a point source, as  expected if the $\rm Ly\alpha$ emission is from extensive star-formation. However, if the  $\rm Ly\alpha$ photons were primarily from an AGN, resonant scattering would be required
to redistribute the $\rm Ly\alpha$ from a point-source to a profile 
reflecting that of the gas (and hence the star-formation). These multiple scatterings of each photon would increase the dust extinction of $\rm Ly\alpha$ above that of the UV continuum, reducing $W(\rm Ly\alpha$), so it might be difficult for even an AGN to produce the observed $\sim 150\rm \AA$.
In contrast, models
(Tenorio-Tagle et al. 1999) and observed sources (e.g. Pettini et al. 1998) suggest that starbursting, with resonant scattering suppressed by outflows, can account for this $W(\rm Ly\alpha)$.

However, some $\rm Ly\alpha$-emitting galaxies do appear to be AGN-dominated.  
Francis, Woodgate and Danks (1997) report 3 galaxies close to a $z=2.38$ QSO, 
with similar  $\rm Ly\alpha$ equivalent widths and velocity widths to CFgA.
These galaxies differ from the CFg in that the $\rm Ly\alpha$ emission is displaced $\sim 10$ kpc from the continuum, and the colours are very red, suggesting old stellar populations. On this basis they were thought to be  obscured AGN rather than
starburst galaxies. 

                                                                                      The high-excitation emission lines, 
CIV($1549\rm \AA$) and HeII($1640\rm\AA$), seen in the MMT spectrum might be interpreted as evidence for an AGN, although the equivalant widths, estimated as  $\sim 12\rm \AA$ and $\sim 6\rm \AA$ respectively,   are smaller than typical of Seyferts. However,  the $\sim 3$ Myr starburst age at which Tenori-Tagle et al. (1999) predict secondary emission from shells corresponds to a similarly brief phase in which starburst galaxies may produce CIV($1549\rm \AA$) and HeII($1640\rm\AA$) in emission
(Leitherer, Roberts and Heckman 1995), due to a peak in the numbers of 
Wolf-Rayet stars. Interestingly,  the $z=2.498$, $W(\rm Ly \alpha \simeq 170\AA)$ starburst galaxy investigated
by Malkan et al. (1996) also showed these emission lines, with similar equivalent widths ($\sim 21$ and $7\rm \AA$ for CIV and HeII), so may
 be another example of this stage of evolution.
  
\subsection{Future Investigations}
It is evident that the Coup Fourr\'e Galaxy, especially its A component, is of special interest in that both its surface brightness evolution and Lyman $\alpha$ emission are at the upper extreme of the class of Lyman break galaxies, or of star-forming galaxies at any redshift, and furthermore it is one of the most luminous galaxies of its type. Our WFPC2 investigations have revealed much about its morphology, but many questions remain unanswered about its nature, and might be addressed as described below:

(i) Further near-infrared spectroscopy of the CFg (Bunker et al. 1995) to detect the redshifted [OII]$3727\rm\AA$,  H$\beta(4861\rm \AA$), [OIII]$                                                                                                                                                                                                                                          5007\rm\AA$ and H$\alpha(6565\rm \AA)$. The equivalent widths of [OII] and H$\alpha$
should give less dust-dependent estimates of the SFR, the 
$\rm H\beta/H\alpha$ ratio may give a direct estimate of the dust extinction,
and
the ratios of the first three lines provide a standard AGN diagnostic (e.g. Tress\'e et al. 1996). The line widths may provide an estimate of the  rotation velocity. The relative velocities of these lines and 
$\rm Ly\alpha$ may indicate gaseous outflows, and should help to explain the high velocity width of $\rm Ly\alpha$ (e.g. Pettini et al. 1998).  We have now obtained these observations using the Keck NIRSPEC.

(ii) Deeper and higher resolution spectroscopy at $\sim 4000\rm\AA$ to reveal the structure of the $\rm Ly\alpha$ line in detail, the strength and profile of the high-excitation emission lines 
CIV($1549\rm \AA$) and HeII($1640\rm\AA$), and other emission or absorption features. The 
spectral index of the UV continuum would provide an improved estimate of the dust extinction (e.g. Calzetti et al. 1995).

(iii) Ideally, a spectrum would be obtained along the CFgA--CFgB axis with sufficient spatial resolution ($<0.4$ arcsec) to  separate the two 
components. This may reveal differences in their star-formation  histories. Spatial resolution of $\rm Ly\alpha$ will provide information on the interaction kinematics, and indicate the contributions of velocity gradient and radial velocity dispersion 
to the broadening of $\rm Ly\alpha$. A velocity gradient of $\Delta\lambda\simeq 7\rm \AA$ -- similar to the line broadening in the CFg -- has already been detected across a lensed $\rm Ly\alpha$ galaxy at an even higher redshift 
(Franx et al. 1997).

(iv) Multi object spectroscopy of the fainter $\rm Ly\alpha$ candidate sources,
to determine the true numbers of strong $\rm Ly\alpha$ emitters, which galaxies are associated with the CFg, and
whether L3:10 is a QSO.  If several of these galaxies are found to be clustered  at the CFg redshift, their velocity dispersion will provide a rough estimate of cluster richness.

(v) Deep imaging of the CFg field in near-infra red $J$ ($1.25\mu \rm m$) and $K$ ($2.2\mu \rm m$) may reveal (or set upper limits on) 
clustering around this galaxy. Galaxies at $2<z<4$ will
 tend to have the reddest $J-K$ colours, and by combining the $J$ and $K$ magnitudes with those from the WFPC2 data it should be possible to derive photometric redshift estimates, and therefore select out all galaxies likely to be at the CFg redshift. In addition, the  $J$ and $K$ magnitudes will provide visible-light absolute magnitudes for the CFg and associated galaxies, 
and hence -- when combined with angular sizes from WFPC2 data -- a reliable measure of their evolution.

\subsection*{Acknowledgements}
Support for this work was provoded by NASA through grant number 
GO-06564.01-95A from the Space telescope Science Institute.

\end{document}